\documentclass[pdftex,onecolumn,epjc3]{svjour3}          

\RequirePackage[T1]{fontenc}

\smartqed  
\usepackage{amsmath}
\usepackage{amssymb}
\usepackage{amsbsy}
\usepackage{amsfonts}
\usepackage{amsopn}
\usepackage{amstext}
\usepackage{graphicx}
\usepackage{amssymb}
\usepackage{amsfonts}
\usepackage{amsmath}
\usepackage{graphicx}
\usepackage[english]{babel}
\usepackage{color}
\usepackage{slashed}
\usepackage{esint}
\usepackage[dvips]{epsfig}
\usepackage[dvips]{graphicx}
\usepackage{float}
\usepackage{units}
\usepackage{textcomp}
\RequirePackage{graphicx}
\RequirePackage{mathptmx}      
\RequirePackage{flushend}
\RequirePackage[numbers,sort&compress]{natbib}
\RequirePackage[colorlinks,citecolor=blue,urlcolor=blue,linkcolor=blue]{hyperref}

\journalname{Eur. Phys. J. C}

\begin{document}

\title{Fermion localization in braneworld teleparallel f(T,B) gravity}

\author{A. R. P. Moreira\thanksref{e1,addr1}
        \and
        J. E. G. Silva\thanksref{addr2}
        \and
        C.A.S. Almeida\thanksref{e2,addr1} 
}

\thankstext{e1}{e-mail:allan.moreira@fisica.ufc.br}
\thankstext{e2}{e-mail:carlos@fisica.ufc.br}

\institute{Universidade Federal do Cear\'a (UFC), Departamento de F\'isica,\\ Campus do Pici, Fortaleza - CE, C.P. 6030, 60455-760 - Brazil\label{addr1}
\and
Universidade Federal do Cariri(UFCA), Av. Tenente Raimundo Rocha, \\ Cidade Universit\'{a}ria, Juazeiro do Norte, Cear\'{a}, CEP 63048-080, Brasil\label{addr2}}

\date{Received: date / Accepted: date}

\maketitle

\begin{abstract}
We study a spin 1/2 fermion in a thick braneworld in the context of teleparallel $f(T,B)$ gravity.  Here, $f(T,B)$  is such that $f_1(T,B)=T+k_1B^{n_1}$ and $f_2(T,B)=B+k_2T^{n_2}$, where  $n_{1,2}$ and $k_{1,2}$ are parameters that control the influence of torsion and the boundary term. We assume Yukawa coupling, where one scalar field is coupled to a Dirac spinor field. We show how the $n_{1,2}$ and $k_{1,2}$ parameters control the width of the massless Kaluza-Klein mode, the breadth of non-normalized massive fermionic modes and the properties of the analogue quantum-potential near the origin.
\end{abstract}

\section{Introduction}

Braneworld scenarios \cite{rs,rs2}, show a new viewpoint of spacetime and enables a new approaches to explain a large number of outstanding issues such as the hierarchy problem \cite{rs2}, the cosmological problem \cite{cosmologicalconstant}, the nature of dark matter \cite{darkmatter} and dark energy. Furthermore, by assuming a warped geometry, the propagation of the gravitational field \cite{Csaki1,Rosa2020} and the gauge field,  \cite{Kehagias}, as well as fermionic fields \cite{Almeida2009}, are governed by the bulk curvature in general relativity (GR). An equivalent theory is the known  teleparallel equivalent of general relativity (TEGR) \cite{Hayashi1979,deAndrade1997,andrade2000,ferraro2007,lobo2012,Aldrovandi,cai2016,koi2020,olmo2020}, that  is constructed in the Weitzenb\"{o}ck manifold, which has vanishing curvature but nonvanishing torsion. Equivalence with GR is provided by the ratio of the torsion $T$ scalar and the Ricci $R$ scalar, which is the boundary term $B$, making TEGR have the same field equations as GR.

The localization mechanism employed for matter fields living in a 5D braneworld scenarios has been the subject of many studies. The study of fermion localization on branes is rich and interesting, yet the most popular method for the localization of fermions is formulated in a rather speculative way \cite{Almeida2009, RandjbarDaemi2000, Liu2009, Liu2009a, Liu2008, Liu2009b, Liu2008b, Obukhov2002,Ulhoa2016}. The same is true in 6D \cite{Dantas2013,Sousa2014,Dantas}. This is because of the freedom one has to propose the Yukawa coupling term. The location of the fermion was studied in several modified gravity models, such as gravity $f(R)$ \cite{Mitra2017, Buyukdag2018, Wang2019} and gravity $f(T)$ \cite{Yang2012,Yang2017}.
 
A new teleparallel gravity model is the $f(T,B)$ gravity, where $B$ is the boundary term \cite{Bahamonde2015, Wright2016, Bahamonde2016}, that has attracted a lot of attention due this model features have,  as well as good agreement with observational data to describe the accelerated expansion of the universe \cite{Franco2020,EscamillaRivera2019}, 
and their significant results  in cosmological perturbations and thermodynamics, and dark energy, and gravitational waves \cite{Bahamonde2016a, Caruana2020,Pourbagher2020,Bahamonde2020a, Azhar2020,Bhattacharjee2020,Abedi2017}. Furthermore, the gravity $ f (T, B) $ was studied in a brane scenario, where it was possible to observe that the additional term $B$ induces changes on the 
energy density causing a split in the brane, also changing the gravitational perturbations \cite{Allan}.

Inspired on the results obtained in \cite{Yang2012, Allan}, we investigate the issue of fermion localization in $f(T,B)$ gravity. In section (\ref{sec1}) we review the main definitions of the teleparallel $f(T,B)$ theory and build the respective braneworld. In the section (\ref{sec2}), we   obtain the solutions of the system and we examined the energy density components of the brane. The section (\ref{sec3}) deals with the fermionic sector of the model using the Yukawa coupling. Finally, additional comments and results are discussed in section (\ref{finalremarks}).

\section{Metric equations}
\label{sec1}

In teleparallel gravity, the vielbein, $h^{\overline{M}}\ _M$ (rather than the metric) are the actual gravitational dynamic variables. We will use the latin letters  ($M , N , Q ,...=0, 1, 2, 3, 4$ ) for the indices related to the bulk, and barred indices ($\overline{M}, \overline{N}, \overline{Q},...=0, 1, 2, 3, 4$ ) for indices related to tangent space. We assume a mostly plus metric signature, i.e., $diag(-1, 1, 1, 1,1)$.

The relevant connection for TEGR is the so-called Weitzenb\"{o}ck connection. An important feature of this connection is that the corresponding spin connection vanishes identically. Thus, the Weitzenb\"{o}ck connection is represented as $\widetilde{\Gamma}^P\ _{NM}=h_{\overline{M}}\ ^P\partial_M h^{\overline{N}}\ _M$, that is, $\nabla_Q h^{\overline{M}}\ _M\equiv \partial_Q h^{\overline{M}}\ _M-  \widetilde{\Gamma}^P\ _{QM} h^{\overline{M}}\ _P =0$,
which is called the condition of absolute parallelism \cite{Aldrovandi}. The Weitzenb\"{o}ck and Levi-Civita connections are related by 
\begin{eqnarray}\label{0.001}
\widetilde{\Gamma}^P\ _{NM}= \Gamma^P\ _{NM} + K^P\ _{NM},
\end{eqnarray}
where $\Gamma^P\ _{NM}$ is the Levi-Civita connection of general relativity, and 
\begin{eqnarray}\label{0.002}
K^P\ _{NM}=( T_N\ ^P\ _M +T_M\ ^P\ _N - T^P\ _{NM})/2
\end{eqnarray}
is defined as the contortion tensor of the Weitzenb\"{o}ck connection \cite{Aldrovandi}.  We take the torsion as $T^{P}\  _{NM}= \widetilde{\Gamma}^P\ _{MN}-\widetilde{\Gamma}^P\ _{NM}$, and we also define a dual torsion tensor, known as a superpotential $S_{P}\ ^{NM}=( K^{NM}\ _{P}-\delta^M_P T^{QN}\ _Q+\delta^N_P T^{QM}\ _Q)/2$ \cite{Aldrovandi}. Therefore, the Lagrangian of TEGR reads
\begin{eqnarray}
\mathcal{L}=- hT/4 ,
\end{eqnarray}
where $T=T_{PMN}S^{PMN}$ is the torsional scalar, $h=\sqrt{-g}$ and $c^4/4\pi G=1$ for simplicity \cite{Aldrovandi}. On the other hand, the Riemann tensor in the Levi-Civita connection
\begin{eqnarray}
R^P\ _{MQN}=\partial_Q\Gamma^P\ _{MN}-\partial_N\Gamma^P\ _{MQ}+\Gamma^P\ _{S Q}\Gamma^S\ _{MN}-\Gamma^P\ _{SN}\Gamma^S\ _{MQ}.
\end{eqnarray}
From the relation between the Weitzenb\"{o}ck connection and the Levi-Civita connection given by Eq.(\ref{0.001}), one can write the Riemann tensor  in the form
\begin{eqnarray}
R^P\ _{MQN}=\nabla_N K^P\ _{MQ}-\nabla_Q K^P\ _{MN}+K^P\ _{SN}K^S\ _{MQ}-K^P\ _{SQ}K^Q\ _{MN},
\end{eqnarray}
whose associated Ricci tensor can then be written as
\begin{eqnarray}
R_{MN}=\nabla_N K^P\ _{MP}-\nabla_P K^P\ _{MN}+K^P\ _{SN}K^S\ _{MP}-K^P\ _{SP}K^S\ _{MN}.
\end{eqnarray}
Using the relations $K^{(MN)S}=T^{M(NS)}=S^{M(NS)}$ and considering that $S^M\ _{PM}=2K^M\ _{PM}=-2T^M\ _{PM}$ along with Eq. (\ref{0.002}), one can get \cite{BLi2010,BLi2011,Sotiriou2010}
\begin{eqnarray}
R _{MN}=-\nabla^P S _{NPM}-g_{MN}\nabla^P T^S\ _{PS}-S^{PS}\ _{N}\ K_{SPN}.
\end{eqnarray}
In turn, Ricci scalar is
\begin{eqnarray}
R=-T-2\nabla^{M}T^{N}\ _{MN}.
\end{eqnarray}
Thus we can identify the boundary term 
\begin{eqnarray}
B\equiv -2\nabla^{M}T^{N}\ _{MN}=\frac{2}{h}\partial_M(h T^M),
\end{eqnarray}
in which $T^M$ is the torsion tensor that can be defined as $T_M=T^N\ _{MN}$.
We can easily see that  GR and TEGR will lead to exactly the same equations \cite{Aldrovandi}. However, this will not be the case if one uses $f(R)$ or $f(T)$ as the Lagrangian of the theory, which therefore corresponds to different gravitational modifications \cite{Abedi2017}.
However, when we consider $f (T, B)$ as the Lagrangian of the theory, we have that $f(T,B)=f(-T+B)$ is the teleparallel equivalent of $f(R)$.

We can consider a modified gravity theory where the gravitational Lagrangian depends on $T$ and $B$ \cite{Abedi2017}. Therefore, we have a gravitational action to $f(T,B)$, namely
\begin{eqnarray}\label{55.5}
\mathcal{S}=-\frac{1}{4}\int h f(T,B)d^5x+\int \mathcal{L}_m d^5x,
\end{eqnarray}
where $\mathcal{L}_m$ is the matter Lagrangian.
We can get the field equations by varying the action with respect to the
vielbein \cite{Abedi2017,Pourbagher2020}
\begin{eqnarray}\label{3.36}
\frac{1}{h}f_T\Big[\partial_Q(h S_N\ ^{MQ})-h\widetilde{\Gamma}^R\ _{SN}S_R\ ^{MS}\Big]+\frac{1}{4}\Big[f-Bf_B\Big]\delta_N^M& &\nonumber\\+\Big[(\partial_Qf_T)+(\partial_Qf_B) \Big]S_N\ ^{MQ} +\frac{1}{2}\Big[\nabla^M\nabla_N f_B-\delta^M_N\Box f_B\Big]&=&-\mathcal{T}_N\ ^M,
\end{eqnarray}
where $\Box\equiv\nabla^M\nabla_M$, $f\equiv f(T,B)$, $f_T\equiv\partial f(T,B)/\partial T$ e $f_{B}\equiv\partial f(T,B)/\partial B$  and $\mathcal{T}_N\ ^M$ is the stress-energy tensor, which in terms of
the matter Lagrangian is given by $\mathcal{T}_a\ ^M=-\delta\mathcal{L}_m/\delta h^a\ _M$. The matter Lagrangian density is taken as
\begin{eqnarray}
\mathcal{L}_m=-h\left[\frac{1}{2}\partial^M\phi\partial_M\phi+V(\phi)\right],
\end{eqnarray}
where $\phi\equiv \phi(y)$ is a background scalar
field that generates the brane.

In our work, we consider the static codimension one braneworld scenario whose metric can be written as
\begin{eqnarray}\label{45.a}
ds^2=e^{2A(y)}\eta_{\mu\nu}dx^\mu dx^\nu+dy^2,
\end{eqnarray}
where  $e^{A(y)}$ is the warped factor. We can choose the vielbein in the form 
\begin{eqnarray}\label{0.658}
h_{\overline{M}}\ ^M=diag(e^A, e^A, e^A, e^A, 1).
\end{eqnarray}
Using the Weitzenbock connection, the torsional scalar and the boundary term are given by
\begin{eqnarray}
 T=-12A'^2	&,& B=-8(A''+4A'^2),	
\end{eqnarray}
where the prime $( ' )$ denotes differentiation with respect to $y$. 

Thus, the gravitational field equations are given as 
\begin{eqnarray}\label{w.0}
\phi''+4A'\phi&=&\frac{d V}{d\phi},\\
\label{w.1}
\frac{1}{4}\Big[f+8(A''+4A'^2)f_B\Big]+6A'^2f_T&=& \frac{1}{2}\phi^2-V,\\
\label{w.2}
12\Big[A'(A'''+8A'A'')(f_{BB}+f_{TB})+3A''A'^2(f_{TT}+f_{BT})\Big]& &\nonumber\\
-\frac{1}{2}(A''+4A'^2)(4f_{B}+3f_{T})+\frac{1}{4}f&=& \frac{1}{2}\phi^2+V.
\end{eqnarray}
We can rewrite equations (\ref{w.1}) and (\ref{w.2}) as
\begin{eqnarray}
6A'^2&=&-\frac{1}{f_T}\Big(P+P_{TB} \Big),\label{0.00005}\\
3A''+12A'^2&=& -\frac{2}{f_T}\Big(\rho+\rho_{TB} \Big),\label{0.00006}
\end{eqnarray}
where
\begin{eqnarray}
 P_{TB}&=&\frac{1}{4}\Big[f+8(A''+4A'^2)f_B\Big],\\
 \rho_{TB}&=& -12\Big[A'(A'''+8A'A'')(f_{BB}+f_{TB})+3A''A'^2(f_{TT}+f_{BT})\Big]\nonumber\\& &+2(A''+4A'^2)f_{B}+\frac{1}{4}f.
\end{eqnarray}
Note that the left side of equations (\ref{0.00005}) and (\ref{0.00006}) is equivalent to that obtained in TEGR. So we can state that modified gravity equations of the motion of the $f(T,B)$ gravities are similar to an inclusion of an additional source with $\rho_{TB}$ and $P_{TB}$.

The diagonal tetrad (\ref{0.658}) represents a good choice among all the possible vielbeins giving metric (\ref{45.a}). In fact, the gravitational field equations do not involve any additional constraints on the function $f(T,B)$ or the scalars $T$ and $B$. Thus, the choice in Eq. (\ref{0.658}) can be regarded as a "good vielbein". Similarly, in the FRW cosmological models the $f(T,B)$ gravitational dynamics preserves the form of the usual Friedmann equations (two equations) \cite{Franco2020,Caruana2020,EscamillaRivera2019, Bahamonde2016a,Pourbagher2020,Bahamonde2020a}. The choice of vielbein is a rather important issue, as it fixes the number of degrees of freedom of the theory, as seen particularly in a gravitational wave analysis of $f(T,B)$ gravity \cite{Capozziello2019}.

\section{Thick brane Solutions}
\label{sec2}

Since the equations (\ref{w.1}) and (\ref{w.2}) form a second-order derivative system, it is difficult to provide an analytical solution for this case. For simplicity, we can take an ansatz \cite{Gremm1999}
\begin{eqnarray}\label{20}
e^{2A(y)}=\cosh^{-2p}(\lambda y),
\end{eqnarray}
where the $p$ parameter modifies the warp variation within the brane core, and $\lambda$ determines the brane width.

We will then propose the cases where $f_1(T,B)=T+k_1B^{n_1}$ and $f_2(T,B)=B+k_2T^{n_2}$ , where $k_{1,2}$ and $n_{1,2}$  are parameters controlling the deviation of the usual teleparallel theory \cite{Allan}.

We follow the approach carried out in Ref. \cite{Allan}, where by manipulating the (\ref{w.1}) and (\ref{w.2}) equations, an equation relating the metric components and the scalar field is obtained. In this case, for $f_1(T,B)$ with equations (\ref{w.1}),(\ref{w.2}) and (\ref{20}), we get \cite{Allan}
\begin{eqnarray}\label{q.5}
\phi'^2(y)=\frac{3}{2}p\lambda^2\mathrm{sech}^2(\lambda y)-\frac{\alpha^{n_1}}{\beta^2}\Big[8^{n_1-1}k_1n_1(n_1-1)(1+4p)\sinh^2(\lambda y)\Big],
\end{eqnarray}
\begin{eqnarray}\label{q.55}
V(\phi(y))&=&\frac{3}{4}\alpha+2^{3n_1-4}k_1(n_1-1)(p\lambda^2)^2\alpha^{n_1-2}\Big\{4 \mathrm{sech}^4(\lambda y)+64p^2\tanh ^4(\lambda y)\nonumber\\ &-&[3n_1+4(8+3n_1)]\mathrm{sech}^2(\lambda y)\tanh ^2(\lambda y)\Big\}.
\end{eqnarray}
We can solve Eq. (\ref{q.5}) to find a function $\phi=g(y)$ that may be inverted to give $y=g^{-1}(\phi)$, which allows us to write the potential in the usual way $V=V(\phi)$.  The thick brane solution for $n_1=1$ is the same obtained in Refs.\cite{Gremm1999,Bazeia2007}, which does not depend on the $k_1$ parameter. For $n_1=2$ we get as a solution the first and second kind elliptic integrals, which depends on parameter $k_1$ \cite{Allan}. 

\begin{figure}
\begin{center}
\begin{tabular}{ccc}
\includegraphics[height=4cm]{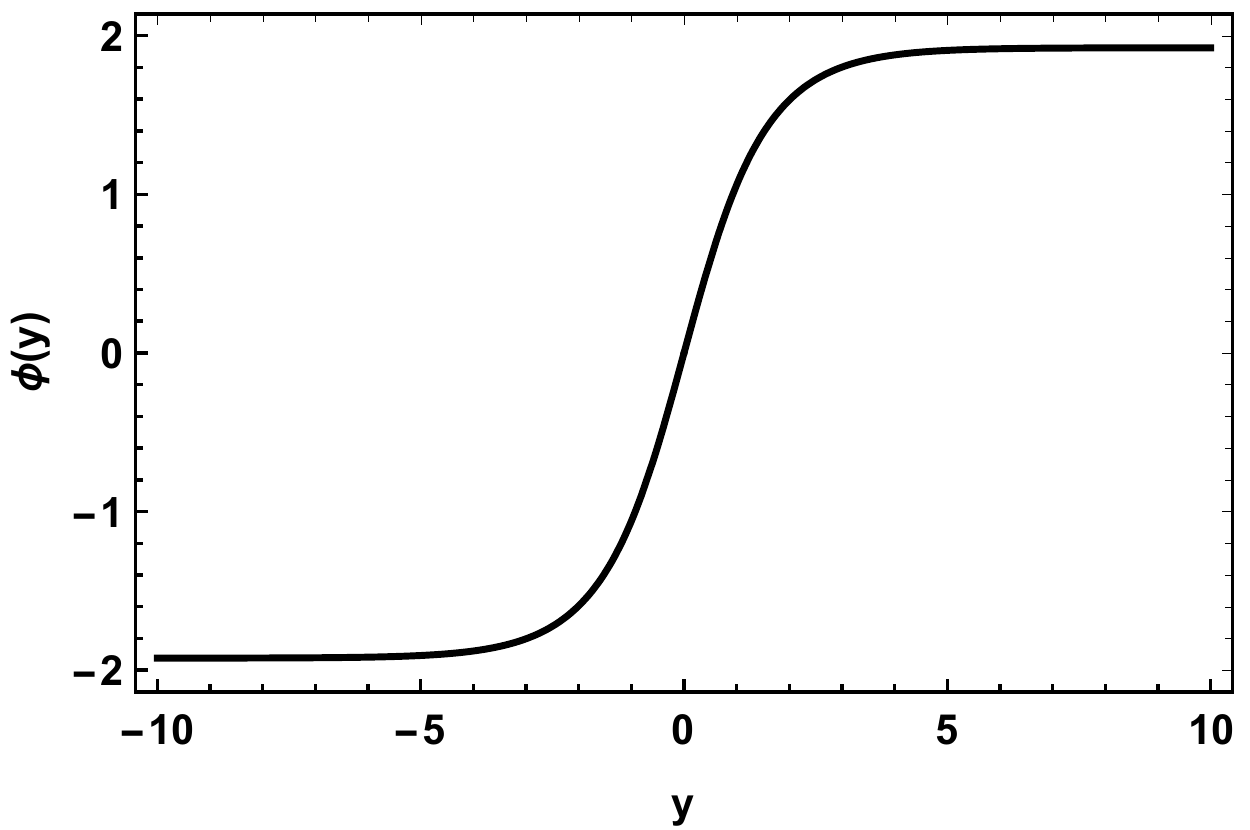} 
\includegraphics[height=4cm]{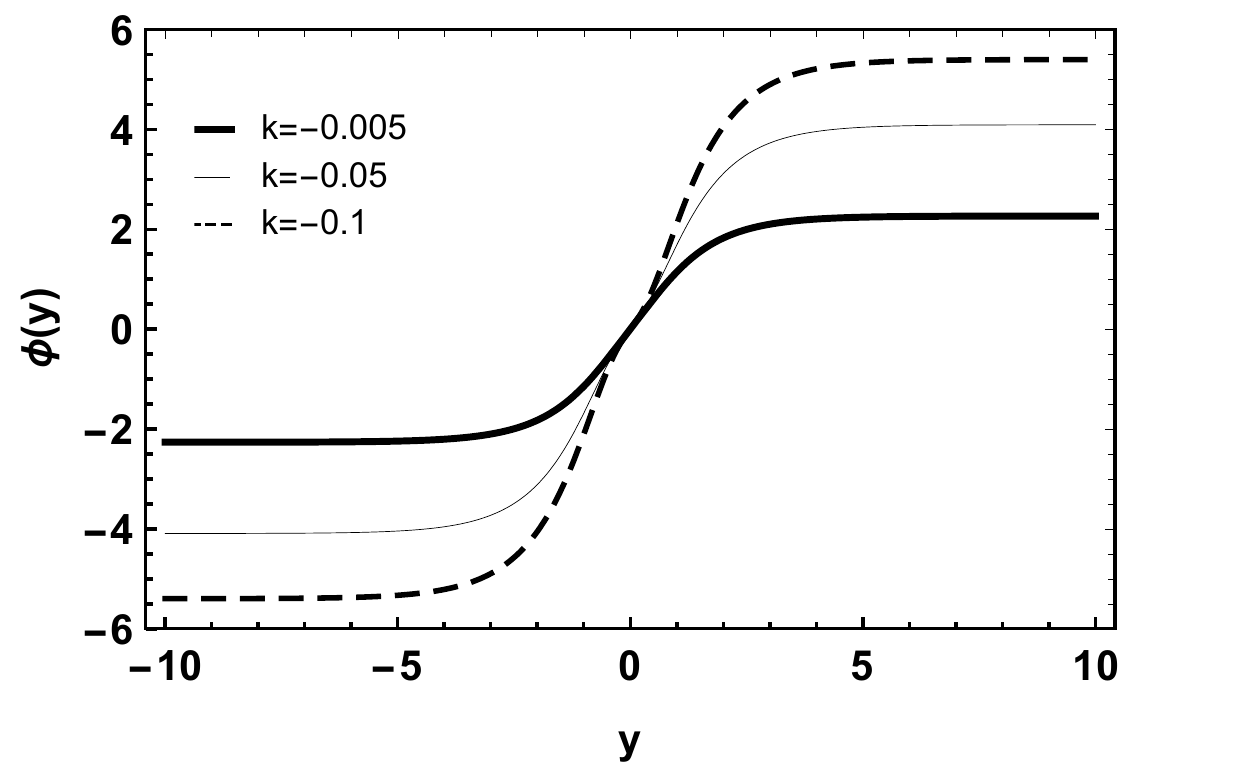}\\
(a) \hspace{6 cm}(b)
\end{tabular}
\end{center}
\caption{ The shape of the scalar $\phi(y)$ for $f_1(T,B)$, where $p=\lambda=1$. (a) for $n_1=1$. (b) for $n_1=2$.
\label{fig1}}
\end{figure}

In Fig. \ref{fig1}, we plotted the $\phi(y)$ field for $f_1(T,B)$.  The $n_1=1$ configuration (figure \ref{fig1} $a$ ) 
 is a kink solution. For $n_1=2$ configuration, for a decreasing value of $k_1$, the solution goes from kink to double-kink, as shown in figure \ref{fig1} ($b$). This feature reflects the brane internal structure, which tends to split the brane. A similar result was obtained in Ref. \cite{Yang2012}. 

For $f_2(T,B)$, we get
\begin{eqnarray}\label{q.6}
\phi'^2(y)=-2^{2n_2-3}(-3)^{n_2}p^{-1}k_2n_2(2n_2-1)\mathrm{csch}^2(\lambda y)[p\lambda\tanh(\lambda y)]^{2n_2},
\end{eqnarray}
\begin{eqnarray}\label{q.606}
V(\phi(y))=4^{n_2-2}(-3)^{n_2}p^{-1}k_2(2n_2-1)[4p-n_2\mathrm{csch}^2(\lambda y)][p\lambda\tanh (\lambda y)]^{2n_2}.
\end{eqnarray}
The setting $ n_2 = 4, 6, ... $ (even numbers) does not present
 a pleasant solution. 
 
\begin{figure}
\begin{center}
\begin{tabular}{ccc}
\includegraphics[height=4cm]{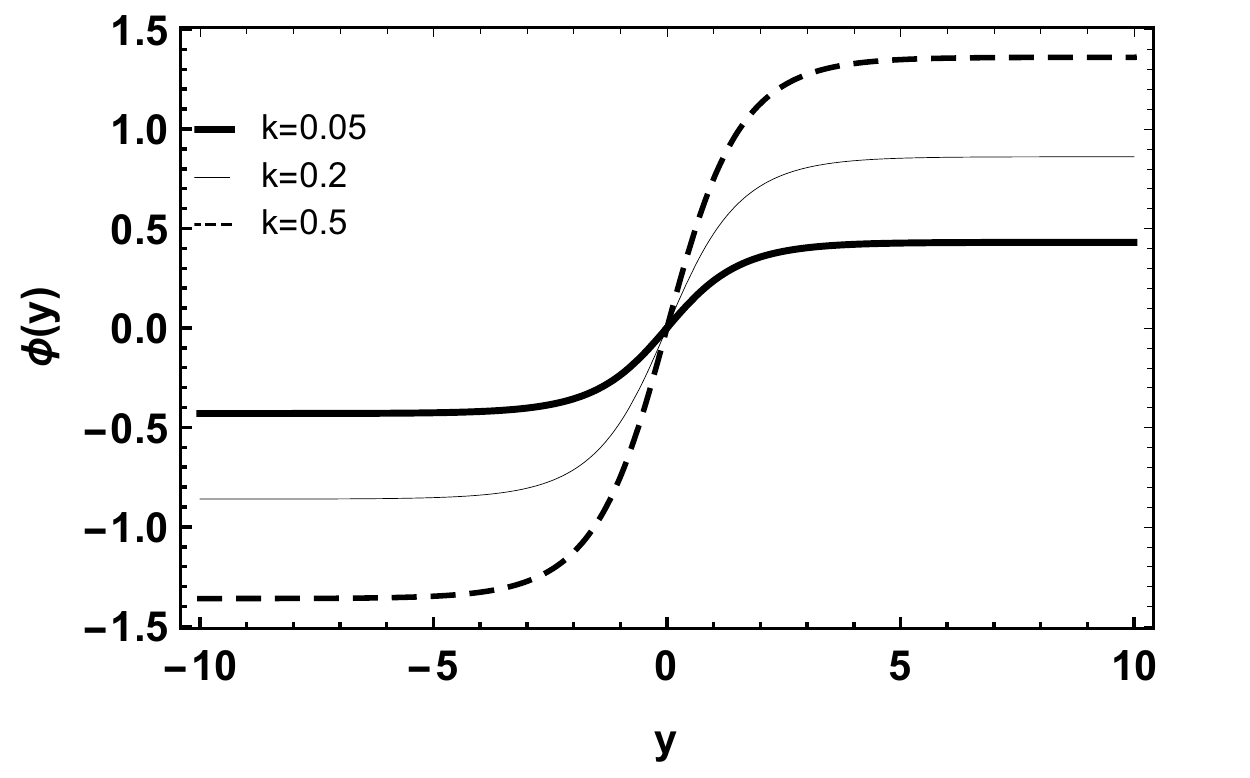} 
\includegraphics[height=4cm]{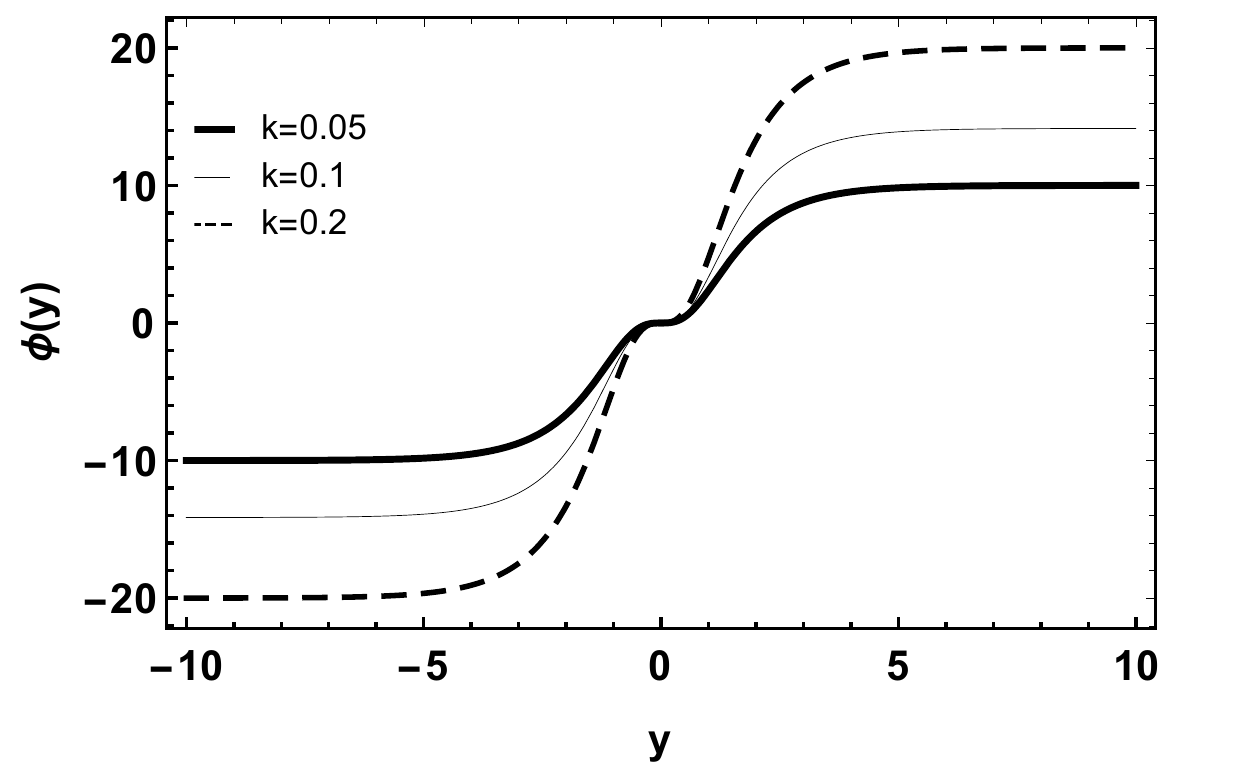}\\
(a) \hspace{6 cm}(b)
\end{tabular}
\end{center}
\caption{ The shape of the scalar $\phi(y)$ for $f_2(T,B)$, where $p=\lambda=1$. (a) for $n_2=1$. (b) for $n_2=3$.
\label{fig2}}
\end{figure}

 In Fig.  \ref{fig2}, we plotted the $\phi(y)$ field for $f_2(T,B)$ varying the parameter $k_2$. For $n_2=1$ configuration  (figure \ref{fig2} $a$ ) 
 we have a kink solution, whereas for  $n_2=3$ configuration we have a double-kink solution (figure \ref{fig2} $b$ ). Again, this feature reflects the brane internal structure, which tends to split the brane.
 
\begin{figure}
\begin{center}
\begin{tabular}{ccc}
\includegraphics[height=4cm]{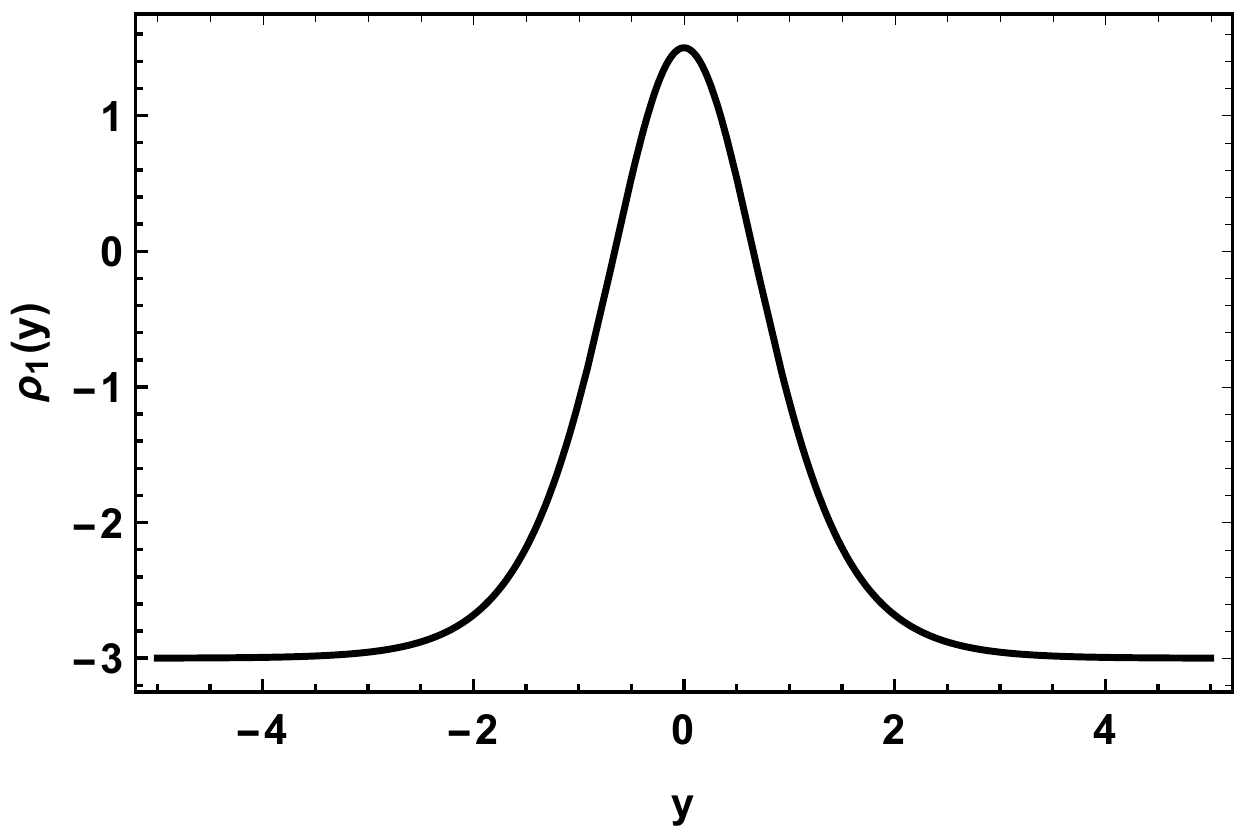} 
\includegraphics[height=4cm]{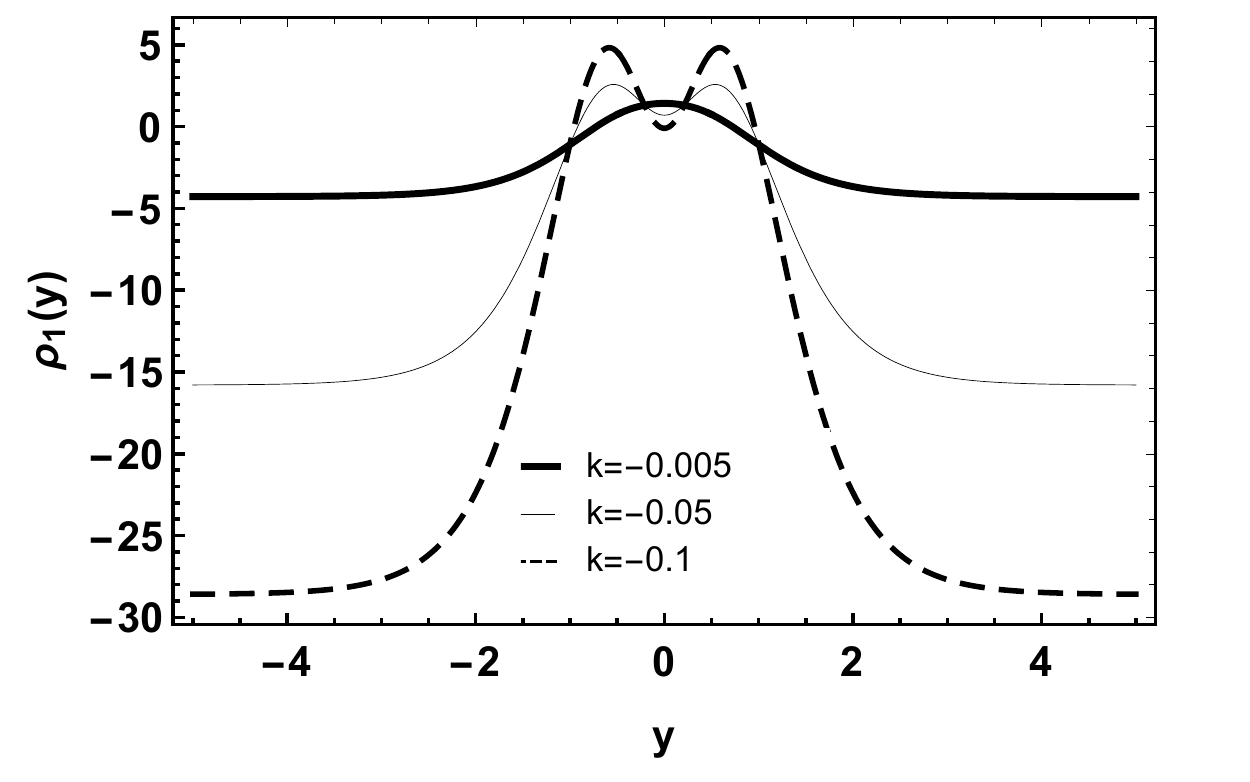}\\
(a) \hspace{6 cm}(b)
\end{tabular}
\end{center}
\caption{ Energy density in the brane for $f_1(T,B)$, where $p=\lambda=1$. (a) for $n_1=1$. (b) for $n_1=2$.
\label{fig3}}
\end{figure}

The energy densities for $f_1(T,B)$ are \cite{Allan}
\begin{eqnarray}
\rho_1(y)&=&\alpha\Bigg[2^{3n_1-2}(n_1-1)k_1 -\frac{8^{n_1-1}(n_1-1)3n_1k_1(1+4p)\sinh^2(\lambda y)}{\beta^2}\Bigg]\nonumber\\&-&3\Big[(p\lambda)^2-\frac{1}{2}(1+2p)p\lambda^2\mathrm{sech}^2(\lambda y)\Big],
\end{eqnarray}
where we defined the functions $\alpha\equiv p \lambda^2[\mathrm{sech}^2(\lambda y)-4p\tanh^2(\lambda y)]$, and $\beta\equiv1+2[1-\cosh(2\lambda y)]p$. In Fig. \ref{fig3}, we plotted the energy densities $\rho_1(y)$ for $f_1(T,B)$, 
with $n_1=1$  (figure \ref{fig3} $a$ ) and  $n_1=2$ (figure \ref{fig3} $b$ ), that includes a new peak varying the parameter $k_1$. 

\begin{figure}
\begin{center}
\begin{tabular}{ccc}
\includegraphics[height=4cm]{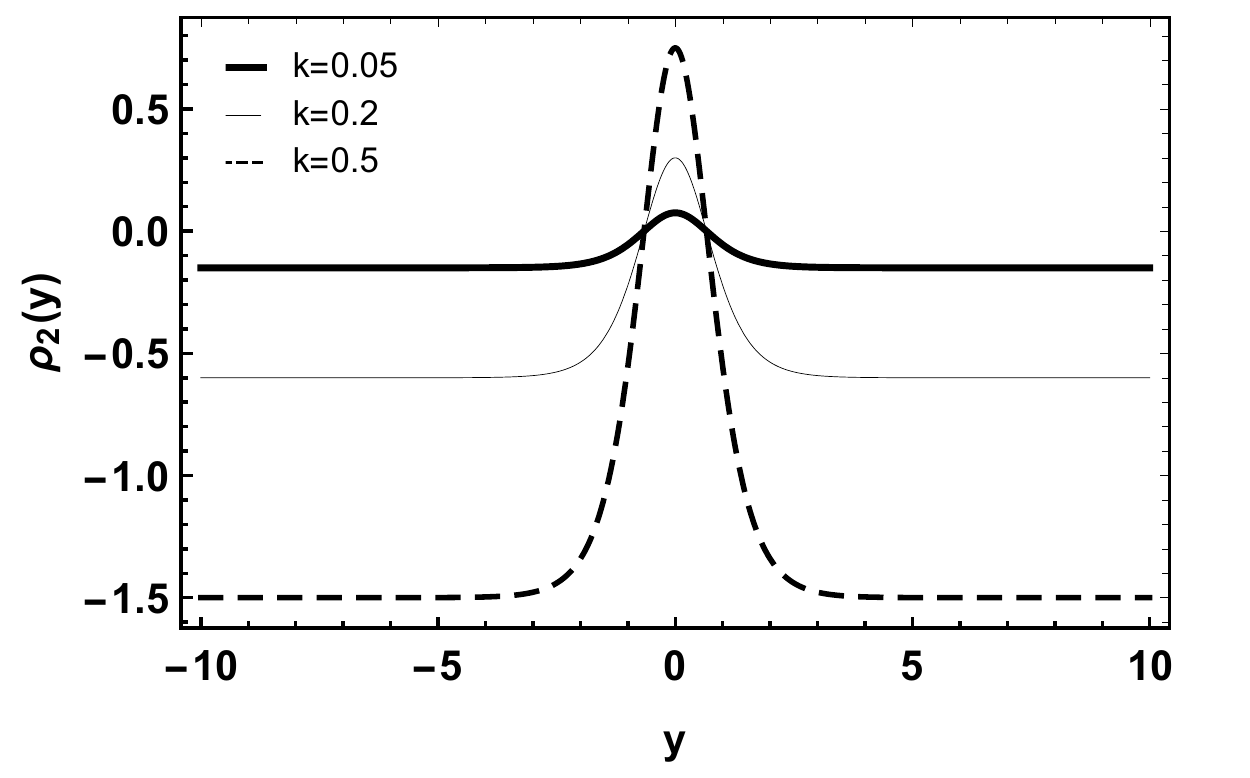} 
\includegraphics[height=4cm]{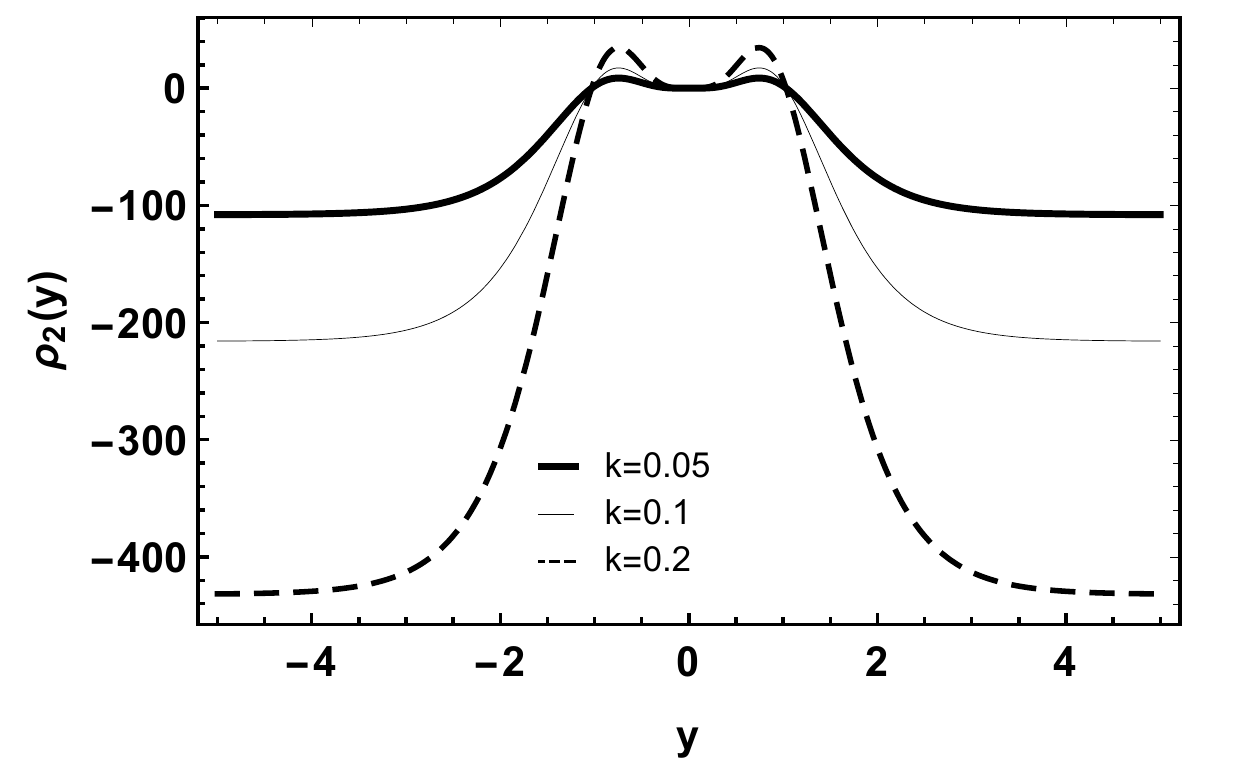}\\
(a) \hspace{6 cm}(b)
\end{tabular}
\end{center}
\caption{ Energy density in the brane for $f_2(T,B)$, where $p=\lambda=1$. (a) for $n_2=1$. (b) for $n_2=3$.
\label{fig4}}
\end{figure}

The energy densities for $f_2(T,B)$ are \cite{Allan}
\begin{eqnarray}
\rho_2(y)=-2^{2n_2-3}(-3)^{n_2}p^{-1}k_2(2n_2-1)\beta\mathrm{csch}^2(\lambda y)[p\lambda\tanh(\lambda y)]^{2n_2}.
\end{eqnarray}
In Fig. \ref{fig4}, we plotted the energy densities $\rho_2(y)$ for $f_2(T,B)$ varying the parameter $k_2$, for $n_2=1$  (figure \ref{fig4} $a$ ) and  $n_2=3$ (figure \ref{fig4} $b$ ), which has two peaks. 

\section{Spin 1/2 Fermions}
\label{sec3}
In this section, we explore the effects of the modified teleparallel $f(T,B)$ on the matter (fermionic) sector.  We changed the variable from $y$ to $z$ in the metric (\ref{45.a}), and so, $dz=e^{-A(y)}dy$ and the metric turns to $ds^2=e^{2A}(\eta^{\mu\nu}dx^\mu dx^\nu+dz^2)$. Considering a well-known Yukawa coupling between the fermion and the scalar field $\phi$, the 5-dimensional Dirac action of a spin $1/2$ fermion minimally coupled to the gravity and to the background scalar $\phi$ is
\begin{eqnarray}\label{1}
\mathcal{S}_{1/2}=\int h \overline{\Psi} \Big(\Gamma^M D_M\Psi -\xi \phi\Psi\Big)d^5x,
\end{eqnarray}
where $\Gamma^M=h_{\overline{M}}\ ^M \Gamma^{\overline{M}}$ are the Dirac curved matrices defined from the Dirac flat matrices $\Gamma^{\overline{M}}$ through the vielbeins. These matrices obey Cliford algebra $\{\Gamma^M,\Gamma^N\}=2g^{MN}$. $D_M$ is the covariant derivative given by $ D_M=\partial_M +\Omega_M$, where \cite{Obukhov2002, Ulhoa2016}
\begin{eqnarray}\label{3}
\Omega_M=\frac{1}{4}\Big(K_M\ ^{{\overline{N}}{\overline{Q}}}\Big)\ \Gamma_{\overline{N}}\Gamma_{\overline{Q}},
\end{eqnarray}
is the spin connection, which for our case is such that $\Omega_\mu=\frac{1}{4}(-\partial_z A)\Gamma_{\mu}\Gamma^{z}$ and $\Omega_z=\partial_z A$.
By choosing the spinor representation \cite{Almeida2009, Dantas, Andrade2001} 
\begin{eqnarray}
\Psi\equiv\Psi(x,z)=\left(\begin{array}{cccccc}
\psi\\
0\\
\end{array}\right),\ 
\Gamma^{\overline{\mu}}=\left(\begin{array}{cccccc}
0&\gamma^{\overline{\mu}}\\
\gamma^{\overline{\mu}}&0\\
\end{array}\right),\ \Gamma^{\overline{z}}=\left(\begin{array}{cccccc}
0&\gamma^4\\
\gamma^4&0\\
\end{array}\right),
\end{eqnarray}
the Dirac equation takes the form
\begin{eqnarray}\label{7}
\Big[\gamma^{\mu}\partial_\mu+\gamma^4\partial_z-\xi e^A\phi\Big]\psi=0.
\end{eqnarray}
We apply a decomposition to the spinor $\psi=\sum_n[\psi_{L,n}(x)\varphi_{L,n}(z)+\psi_{R,n}(x)\varphi_{R,n}(z)]$, being $\gamma^4\psi_{R,L}=\pm\psi_{R,L}$ e $\gamma^\mu\partial_\mu\psi_{R,L}=m\psi_{L,R}$. So, we have the coupled equations
\begin{eqnarray}\label{9}
\Big[\partial_z+\xi e^A \phi\Big]\varphi_{L}(z)=m \varphi_{R}(z),\nonumber\\
\Big[\partial_z-\xi e^A\phi\Big]\varphi_{R}(z)=m \varphi_{L}(z).
\end{eqnarray}
These equations can be decoupled and reduced to Schroëdinger-like equations
\begin{eqnarray}\label{10}
\Big[-\partial^2_z+V_L(z)\Big]\varphi_{L}(z)=m^2 \varphi_{L}(z),\nonumber\\
\Big[-\partial^2_z+V_R(z)\Big]\varphi_{R}(z)=m^2 \varphi_{R}(z),
\end{eqnarray}
where
\begin{eqnarray}\label{11}
V_L(z)=U^2 -\partial_{z}U,\nonumber\\
V_R(z)=U^2 +\partial_{z}U,
\end{eqnarray}
and $U=\xi e^A \phi$ is the so-called superpotential. The supersymmetric structure of the potentials \eqref{11} leads to a massless mode in the form

\begin{eqnarray}
\varphi_{R0,L0}(z)\propto \exp{\Bigg[\pm\int\xi \phi e^{A}dz\Bigg]},
\end{eqnarray}
where $\phi e^{A}|_{z\rightarrow\pm\infty}\rightarrow0$. In other words, the zero mode for fermions can be localized on the
brane for positive $\xi$ \cite{Yang2012}. 

\begin{figure}
\begin{center}
\begin{tabular}{ccc}
\includegraphics[height=4cm]{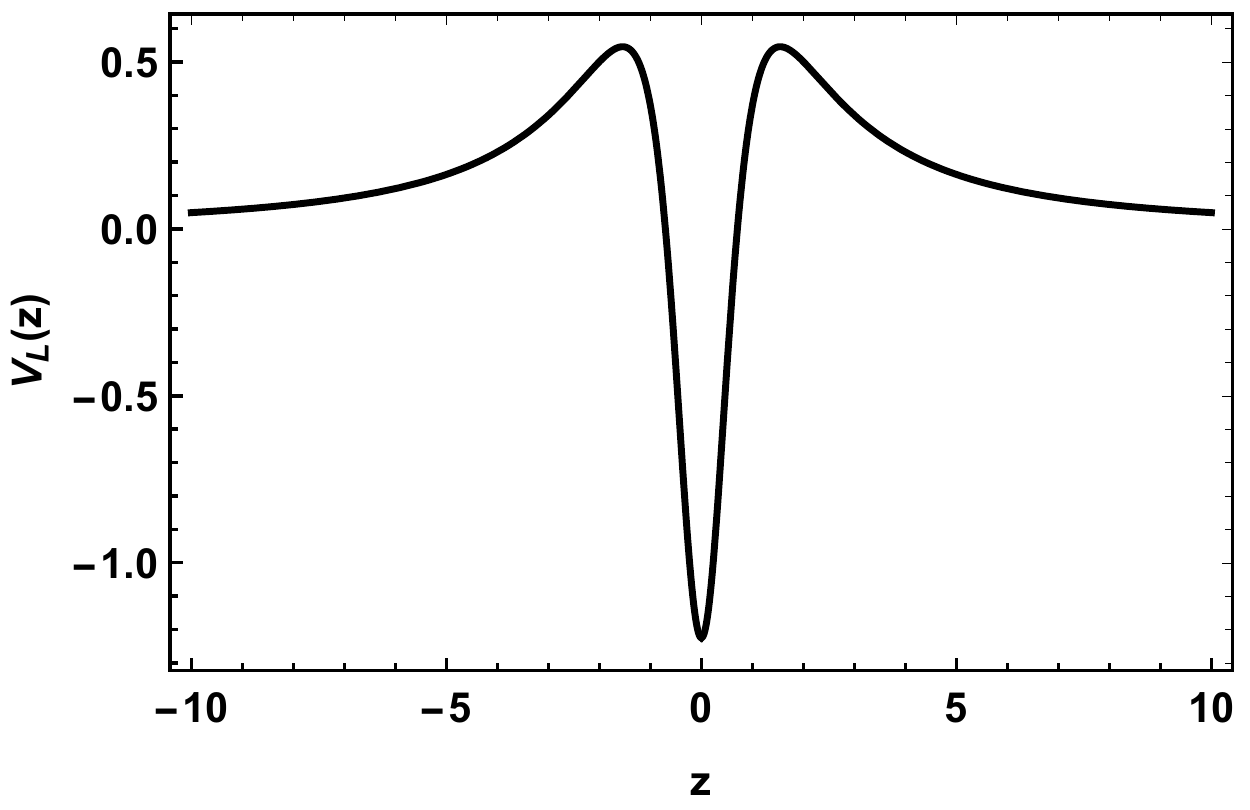} 
\includegraphics[height=4cm]{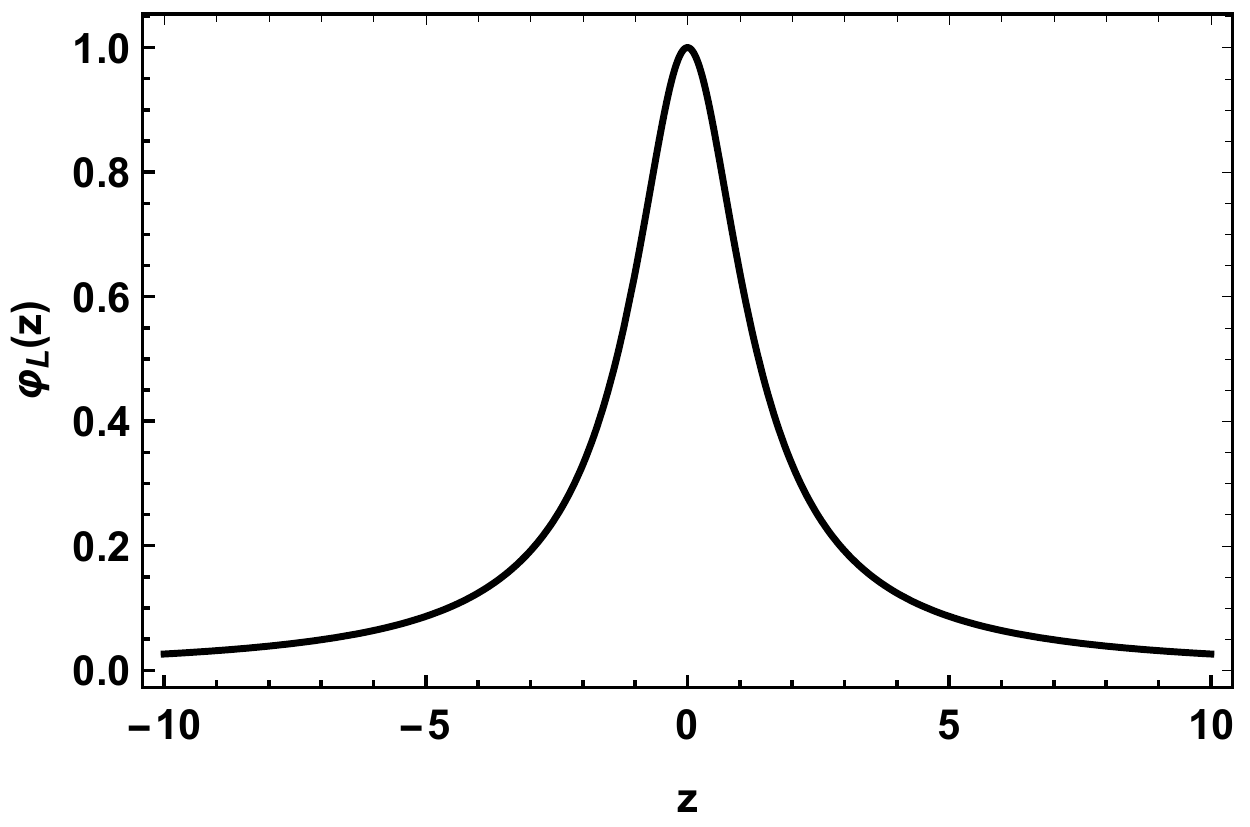}\\
(a) \hspace{6 cm}(b)\\
\includegraphics[height=4cm]{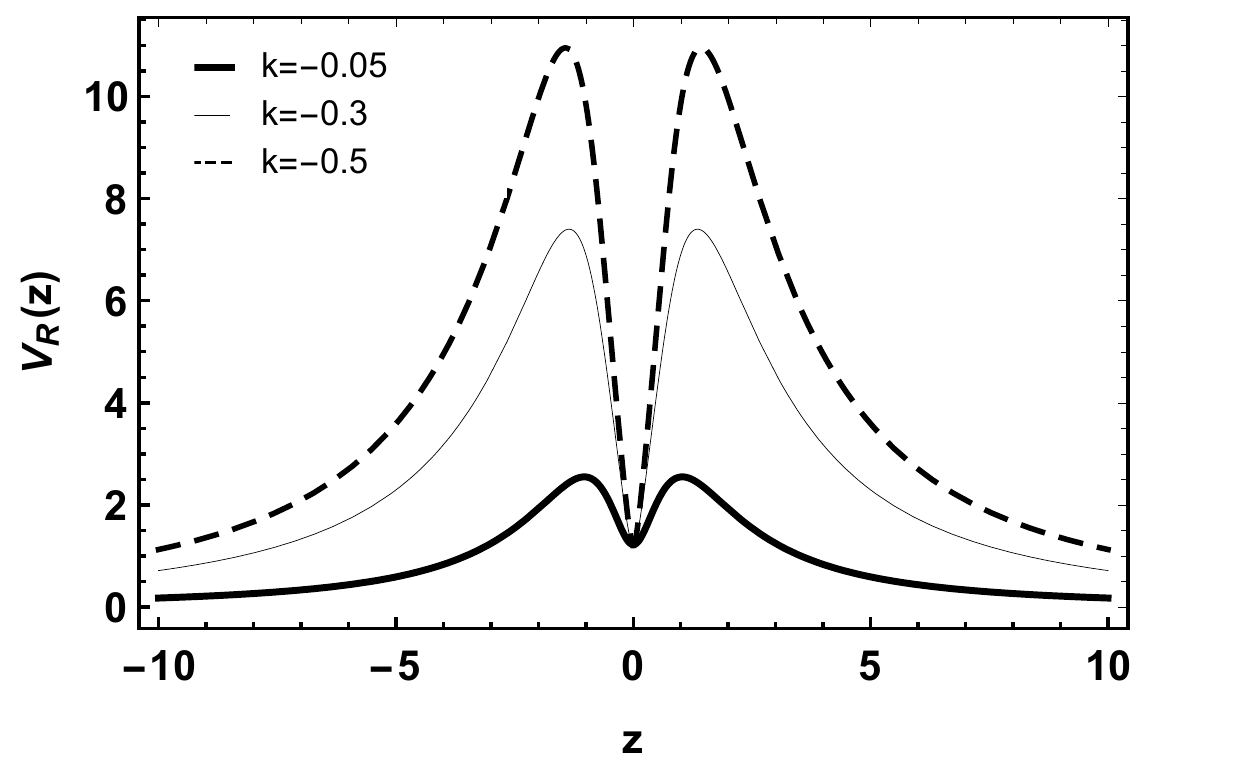} 
\includegraphics[height=4cm]{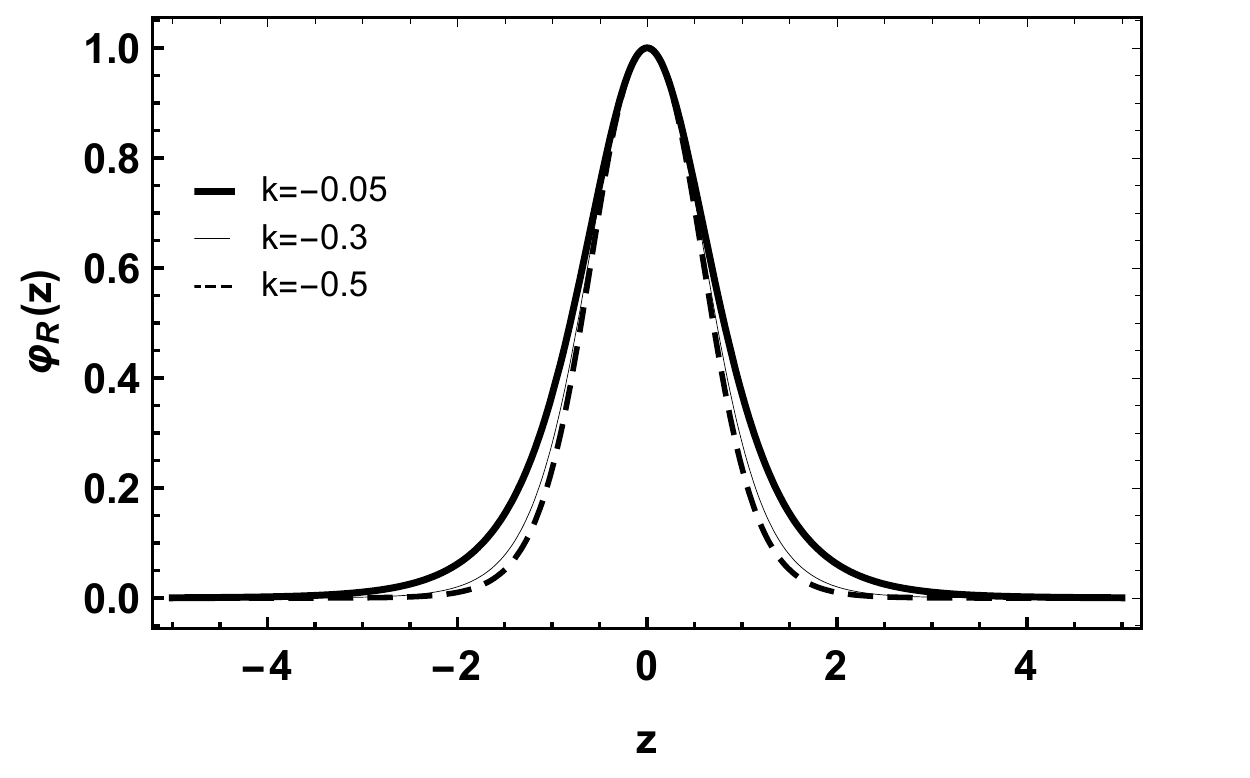}\\
(c) \hspace{6 cm}(d)
\end{tabular}
\end{center}
\caption{$V_L$ (a) and $\varphi_{L}$ (b) for  $f_1(T,B)$  with $n_1=1$. $V_R$ (c) and $\varphi_{R}$ (d)  for $f_1(T,B)$ with $n_1=2$ ($p=\lambda=\xi=1$). 
\label{fig5}}
\end{figure}

\begin{figure}
\begin{center}
\begin{tabular}{ccc}
\includegraphics[height=4cm]{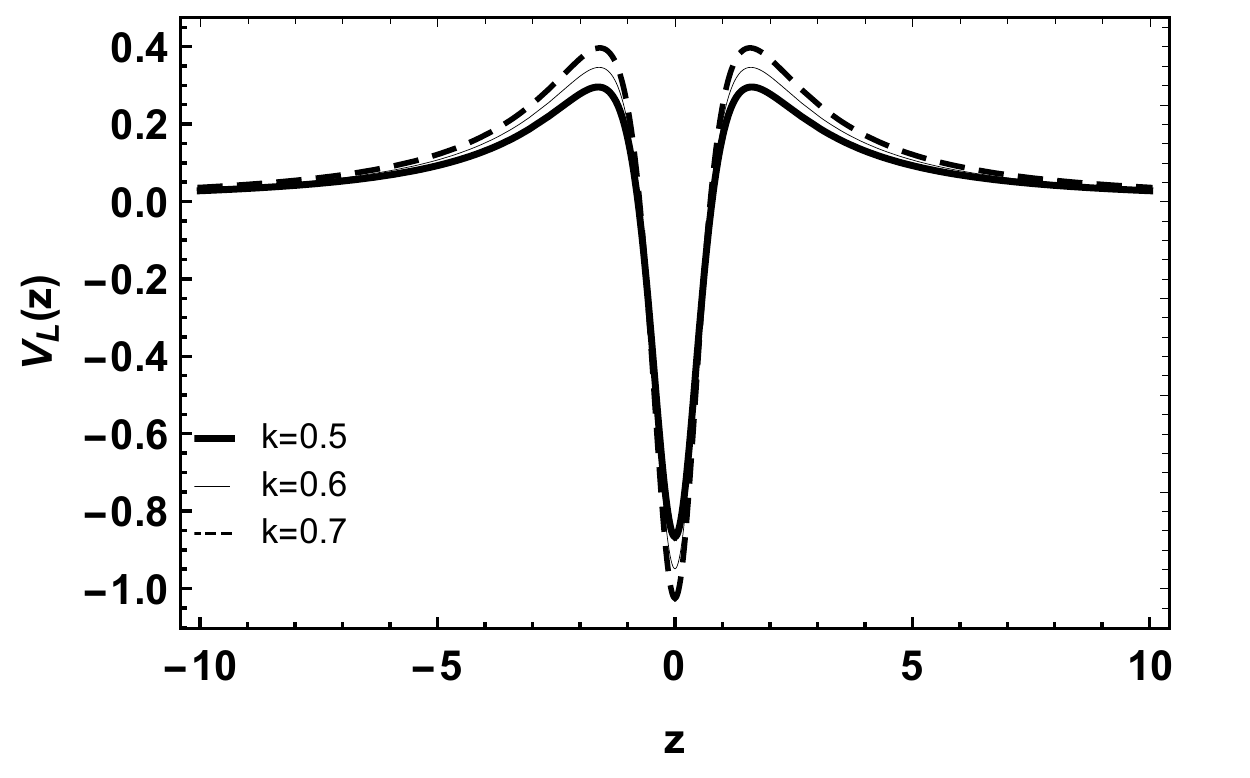} 
\includegraphics[height=4cm]{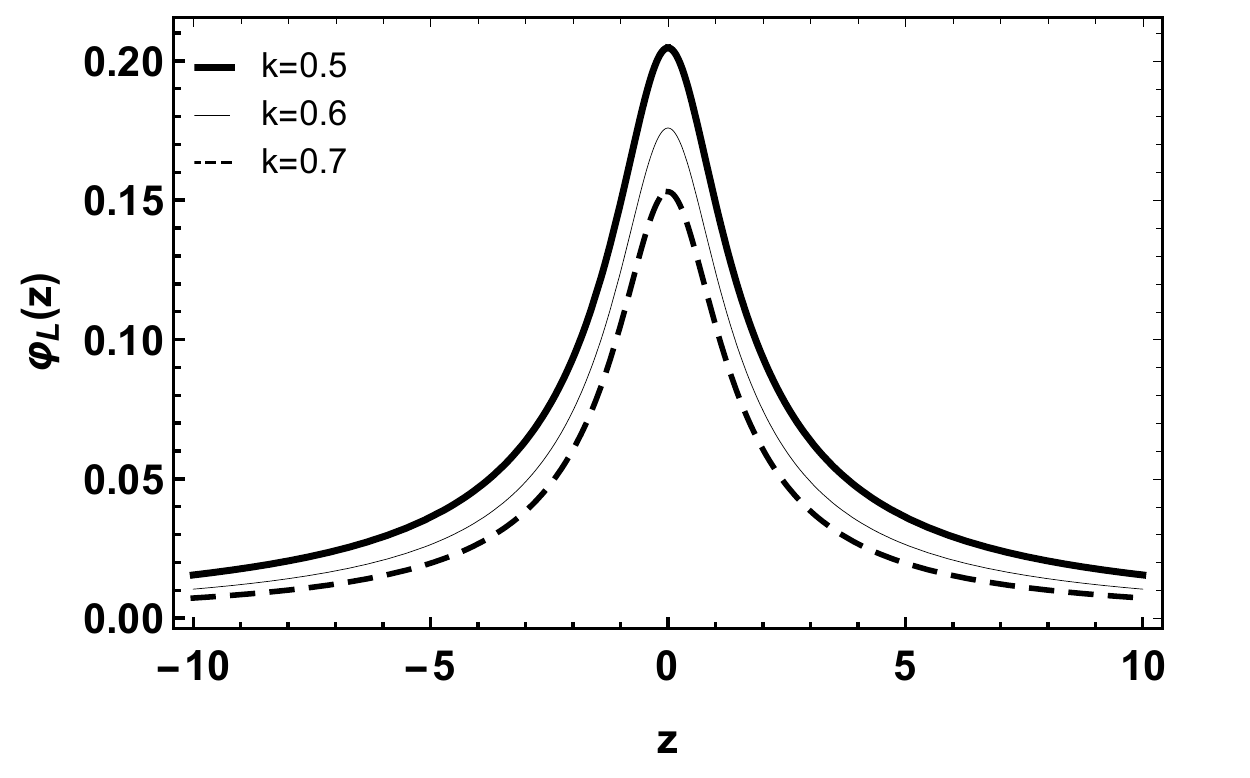}\\
(a) \hspace{6 cm}(b)\\
\includegraphics[height=4cm]{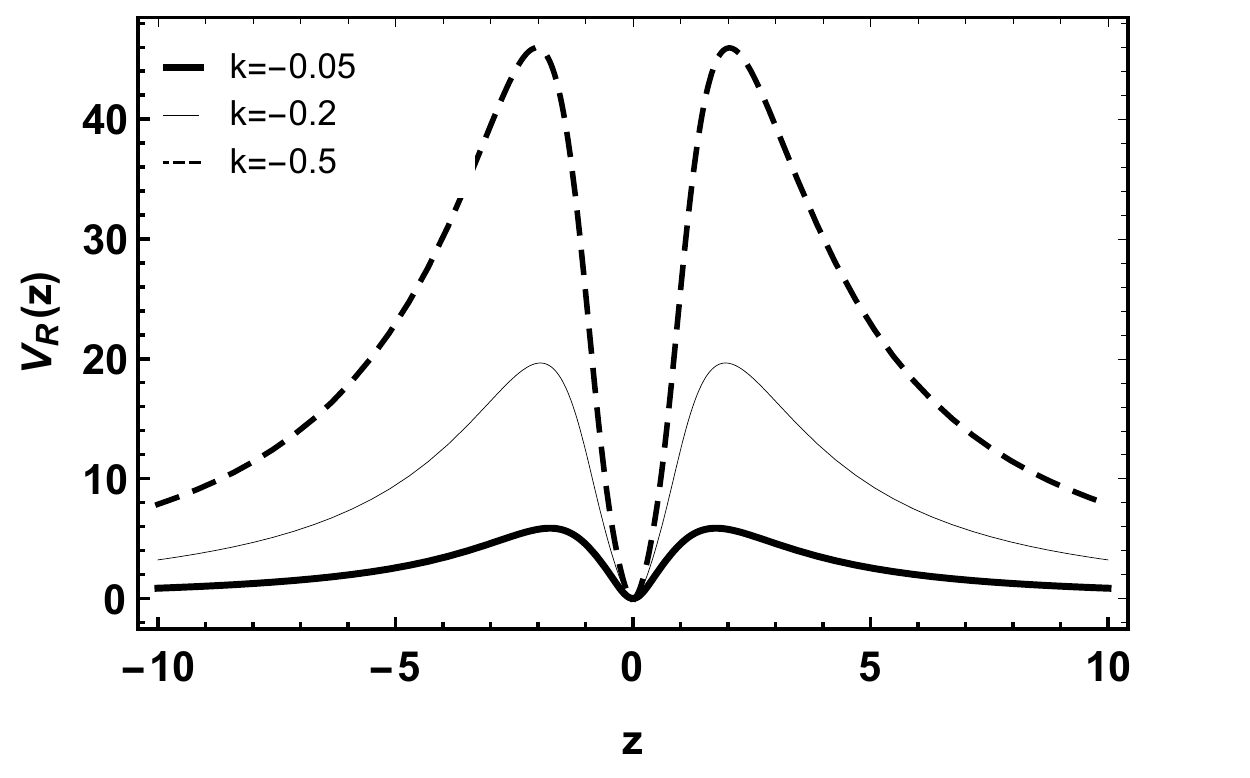} 
\includegraphics[height=4cm]{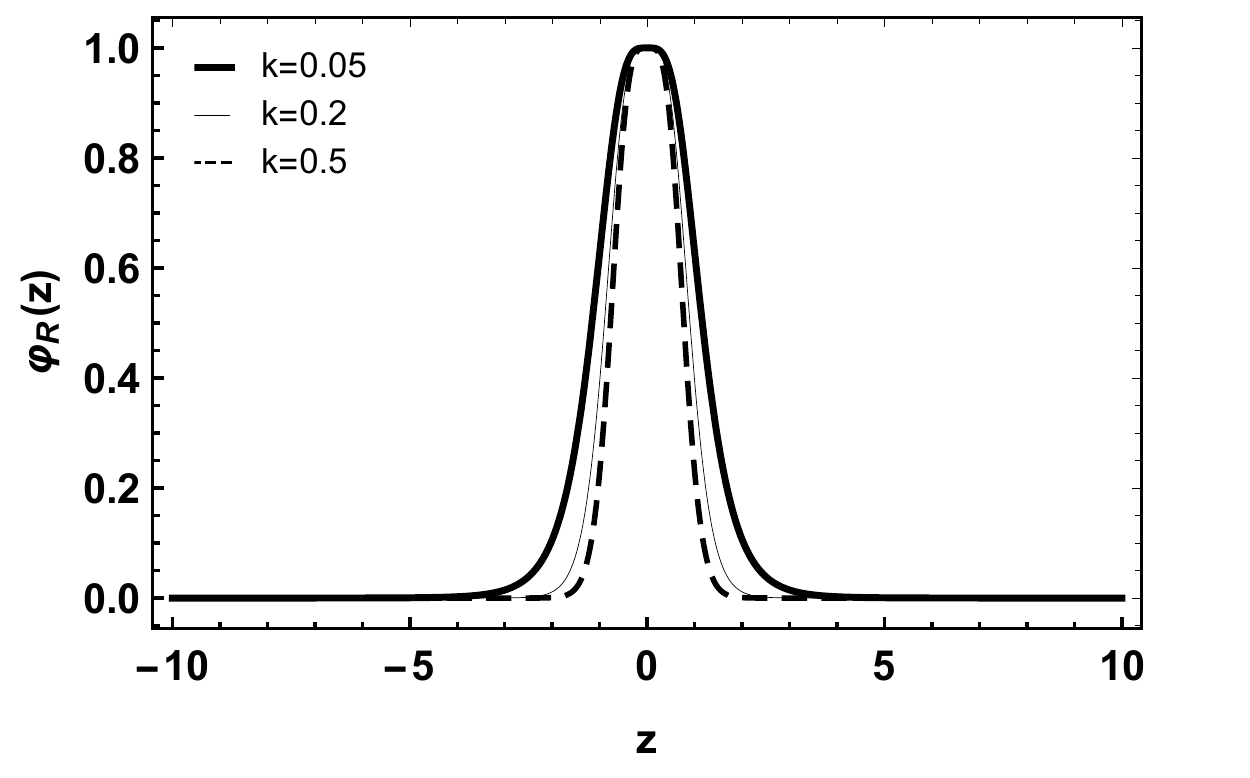}\\
(c) \hspace{6 cm}(d)
\end{tabular}
\end{center}
\caption{$V_L$ (a) and $\varphi_{L}$ (b) for  $f_2(T,B)$  with $n_2=1$. $V_R$ (c) and $\varphi_{R}$ (d)  for $f_2(T,B)$ with $n_2=3$ ($p=\lambda=\xi=1$). 
\label{fig6}}
\end{figure}

For $f_1(T,B)$ with $n_1=1$ only left-chiral fermions can be localized on the brane. For $n_1=2$ only right-chiral fermions can be localized on the brane. In this case note that the smaller the $k_1$ parameter, the more localized the mode becomes, as can be seen in the figure (\ref{fig5}). For $f_2(T,B)$ with $n_2=1$ only left-chiral fermions can be localized on the brane. Now, the higher the parameter $ k_1 $, the more localized the mode becomes. On the other hand, for $n_1=3$ only right-chiral fermions can be localized on the brane, and so, the smaller the $k_1$ parameter, the more localized the mode becomes, as can be seen in the figure (\ref{fig6}).

Note that the effective potentials are even functions, so the wave functions will be either even or odd. We can analyze numerically the  Eq.(\ref{10}). For that we impose the following conditions: $\varphi_{even}(0)=c$, $\partial_z\varphi_{even}(0)=0$, $\varphi_{odd}(0)=c$, and $\partial_z\varphi_{odd}(0)=0$, where $c$ is a constant\cite{Almeida2009,Liu2009,Liu2009a}. Here $\varphi_{even}$ and $\varphi_{odd}$ denote the even and odd parity modes of $\varphi_{R,L}(z)$, respectively. 

\begin{figure}
\begin{center}
\begin{tabular}{ccc}
\includegraphics[height=4cm]{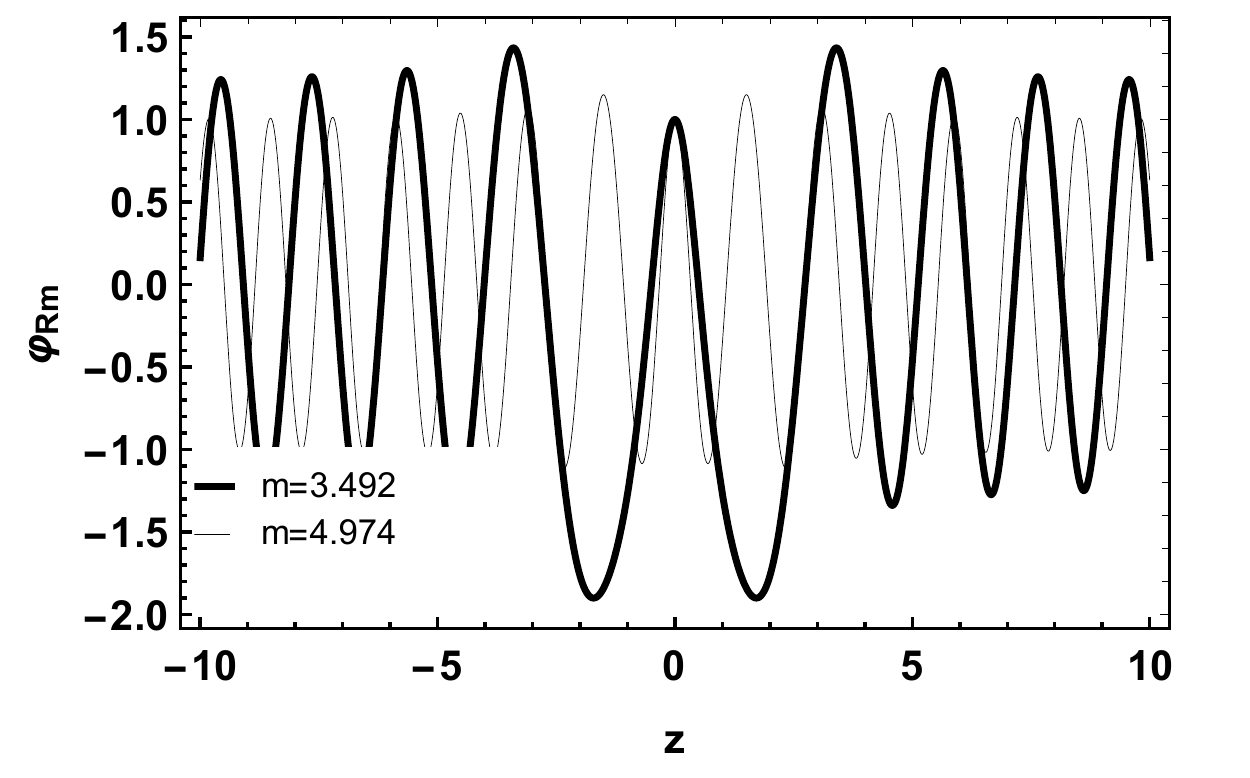} 
\includegraphics[height=4cm]{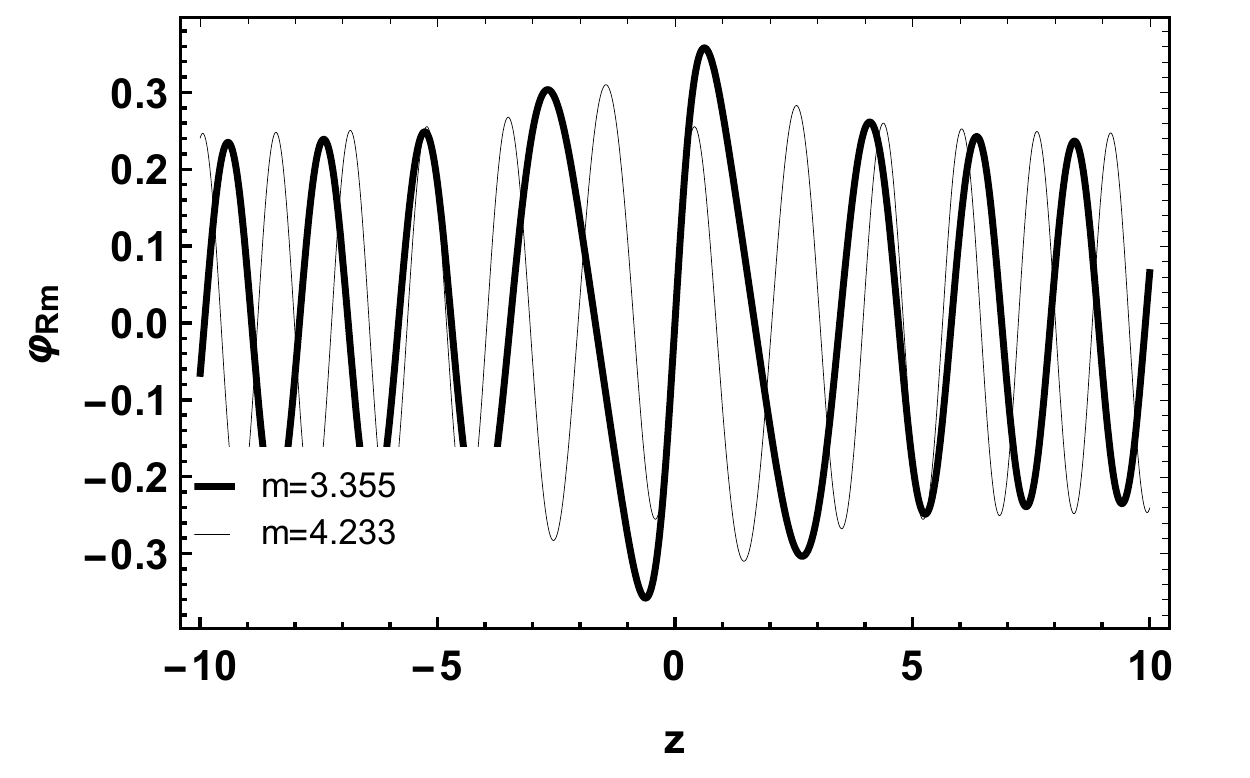}\\
(a) \hspace{6 cm}(b)\\
\includegraphics[height=4cm]{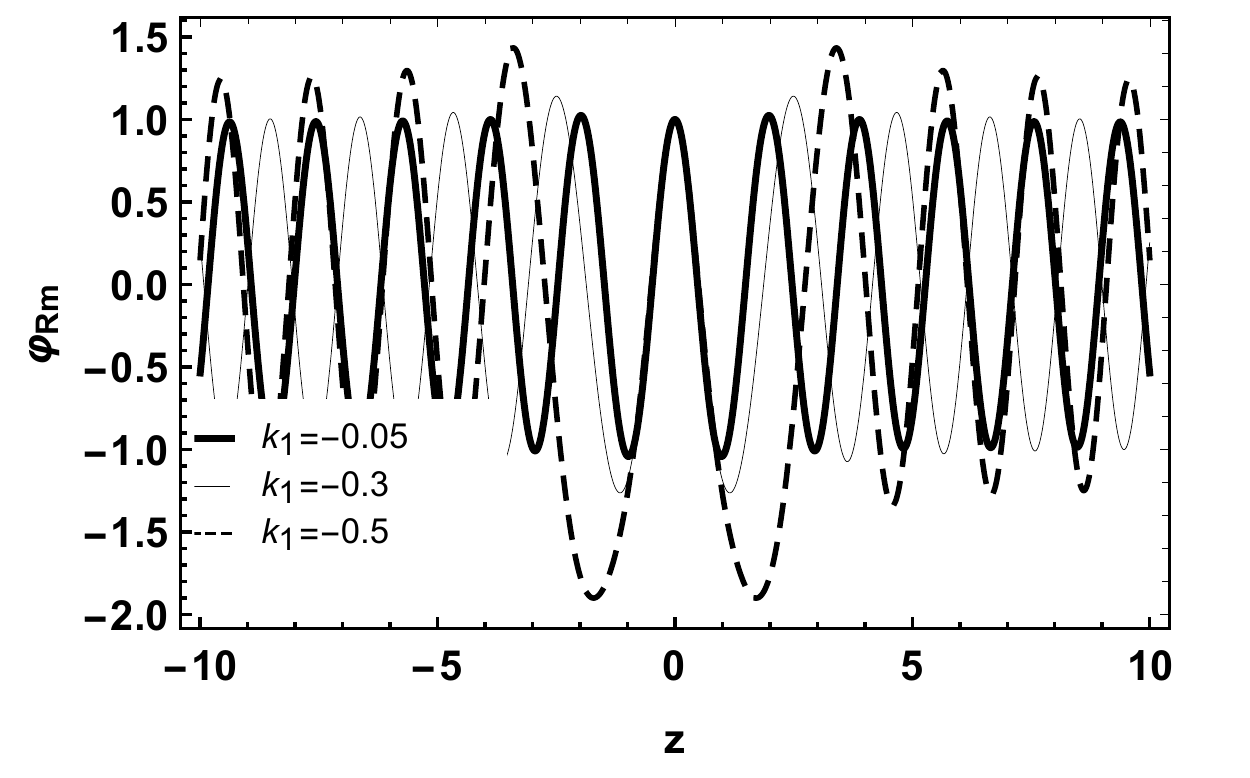} 
\includegraphics[height=4cm]{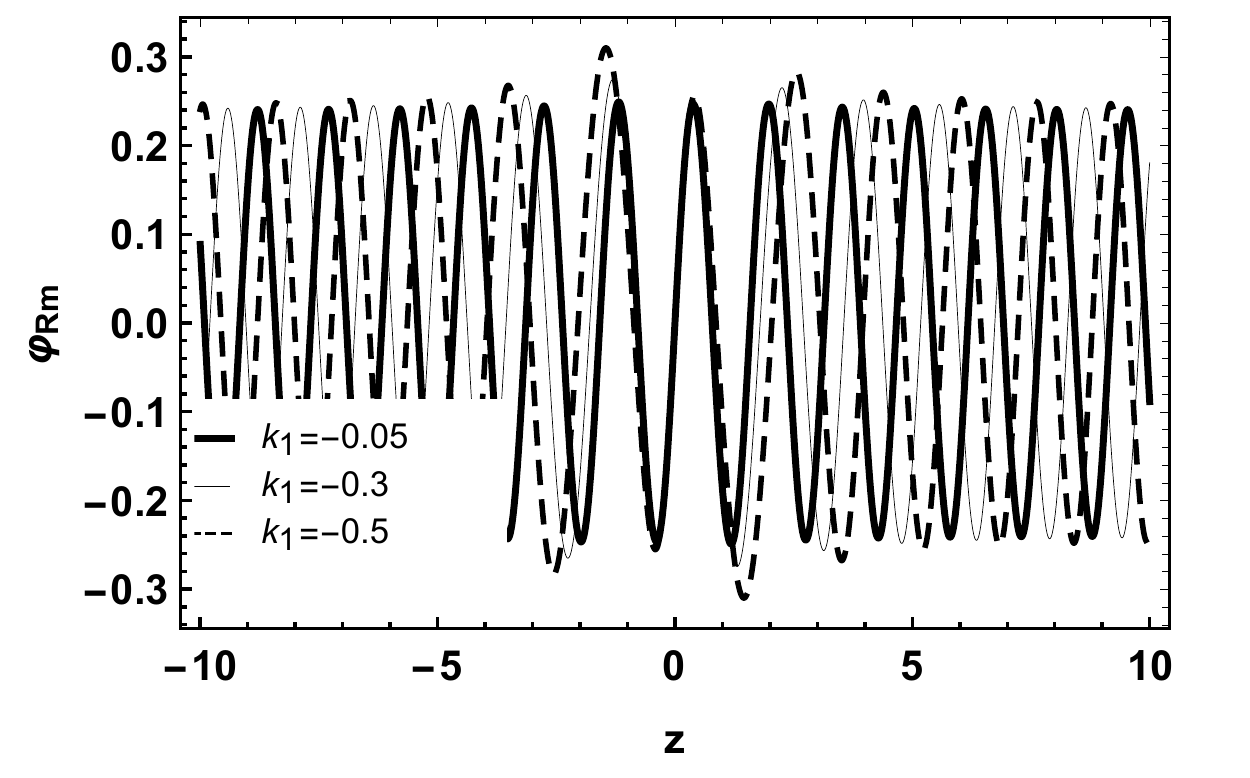}\\
(c) \hspace{6 cm}(d)\\
\includegraphics[height=4cm]{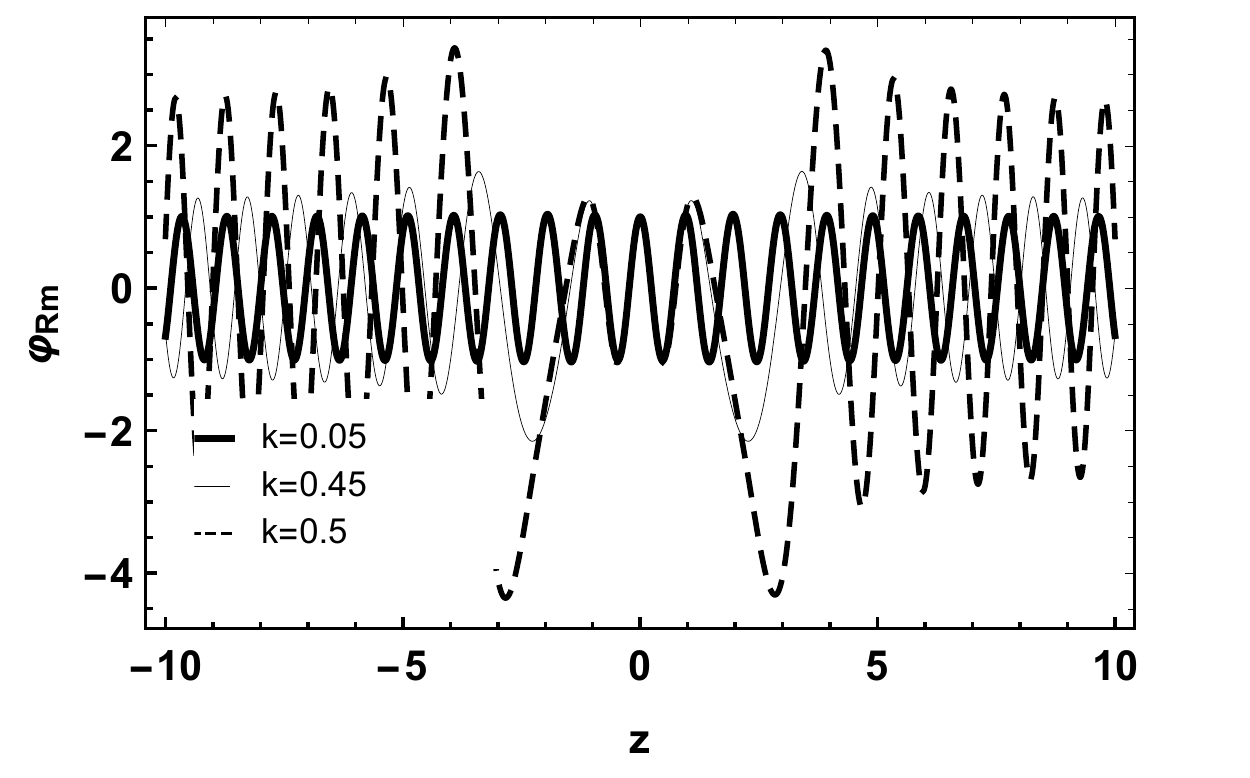} 
\includegraphics[height=4cm]{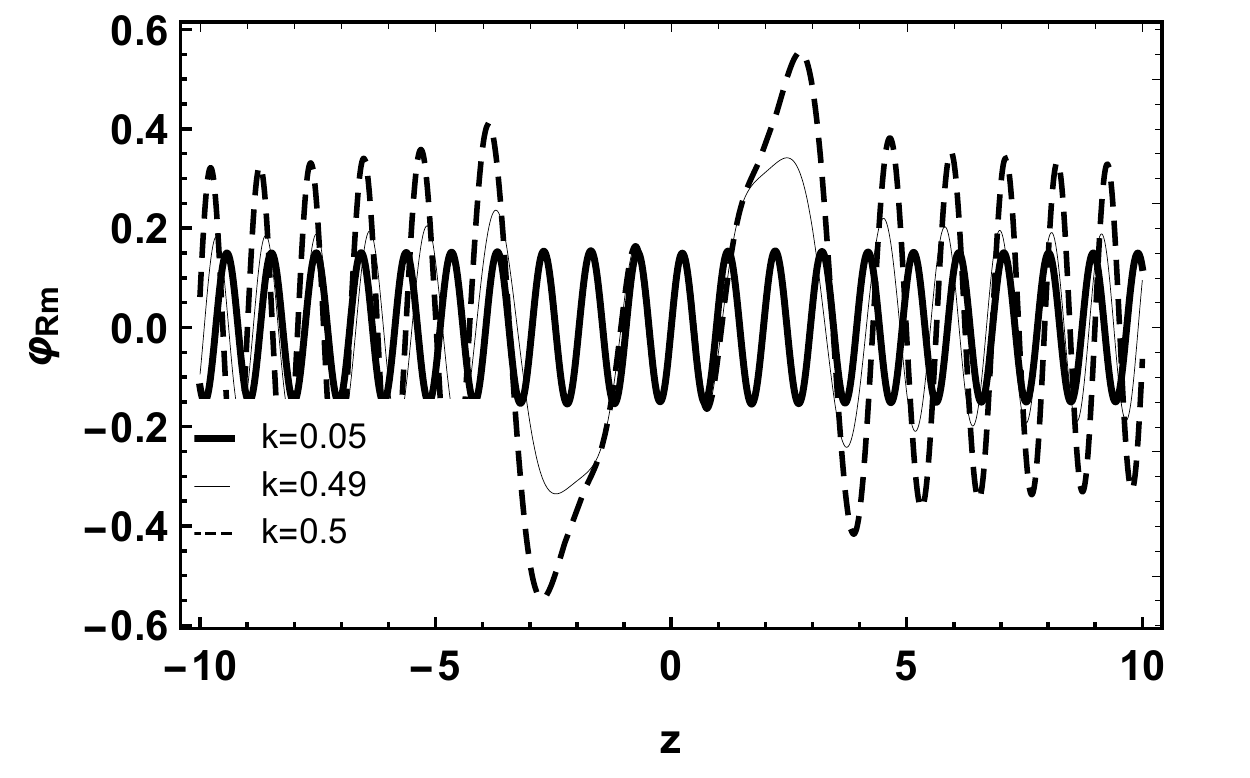}\\
(e) \hspace{6 cm}(f)
\end{tabular}
\end{center}

\caption{Non-normalized  massive fermionic modes for $f_1(T,B)$, with $n_1=2$ and $k_1=-0.5$. For $\varphi_{even}$ (a) and $\varphi_{odd}$ (b). By varying $k_1$, $\varphi_{even}$ with $m=3.492$ (c) and $\varphi_{odd}$ with $m=4.233$ (d). For $f_2(T,B)$ with $n_2=3$, $\varphi_{even}$ with $m=6.72$ (e) and $\varphi_{odd}$ with $m=6.711$  (f) ($p=\lambda=\xi=1$).}
\label{fig7}
\end{figure}
 
 As depicted in Fig. \ref{fig7}, the asymptotic divergence of the massive modes shows that they form non-localized states,
which is a behavior typical of plane wave oscillations, characteristic of a free mode. This shows that these modes represent massive fermions that certainly will be leaked from the brane. 

For $f_1(T,B)$, both for $n_1=1$ and $n_1=2$, the greater the mass, the more oscillations we obtain, as can be seen in the figure \ref{fig7} ($a$ and $b$) for $n_1=2$. In figure \ref{fig7} ($c$ and $d$), we observe that when decreasing the value of $k_1$, the greater the amplitude of the oscillation, mainly near the brane. For $f_2(T,B)$, both for $n_2=1$ and $n_2=3$, the greater the mass, the more oscillations we obtain. Increasing the value of $k_2$, greater the amplitude of the oscillation, mainly near the brane as we can see in the figure \ref{fig7} ($e$ and $f$) for $n_2=3$.

\section{Final remarks}
\label{finalremarks}

In this work we considered a braneworld in the context of the $f(T,B)$ modified teleparallel gravity constructed with one scalar field.  We propose two particular cases for $ f(T,B)$, namely $f_1(T,B)=T+k_1B^{n_1}$ and $f_2(T,B)=B+k_2T^{n_2}$.  In both cases the torsion and  boundary term produce an inner brane structure tending to split the brane. We also find that the  $n_{1,2}$ and $k_{1,2}$ parameters determines whether the domain wall solution is a kink or double-kink. For $f_1(T,B)$ where $n_1=2$, with the decreasing of the contribution of $k_1$, the configuration of the solution changes from a kink to double-kink. The same is true for $f_2(T,B)$  where $n_2=3$, when we increase the value of parameter $k_2$. The thick brane undergoes a phase transition evinced by the energy density components. Similar behavior was found for $f(T)$ in Ref \cite{Yang2012}. 

We considered a simple Yukawa coupling between the scalar and the spinor field. We notice that potentials feel the division of brane when we vary $n_{1,2}$ and $k_{1,2}$, the same happens with the zero modes, which become more localized. We note that for $f_1(T,B)$ where $n_1=1$, only left-chiral fermions are located, the same is true for $f_2(T,B)$ with $n_2=1$. For $f_1(T,B)$ where $n_1=2$, only right-chiral fermions are located, the same is true for $f_2(T,B)$ with $n_2=3$. The massive fermionic modes are dependent on the parameters that control torsion and boundary term. This is well evidenced for $f_1(T,B)$ with $ n_1 = 3 $ since that decreasing the value of $k_1$, increases the amplitude of the ripples making them more intense and presenting ripples within the brane. The same goes for $f_2(T,B)$ when increasing the value of $k_2$, which is very evident for $n_2=3$. Therefore, the brane splitting process leads to modifications of the massive fermionic modes inside the thick brane. The interaction of the massive modes with the torsion and boundary term is more intense in the brane core where the amplitude and the rate of growth depend on the parameters $n_{1,2}$ and $k_{1,2}$.

Although only one massless chiral mode was found for each configuration $n_{1,2}$ and $k_{1,2}$, only left-handed spin $1/2$ fermions were detected so far. The configurations where the right-handed massless mode is localized on the brane are beyond the standard model states. The absence of left-handed massless mode can be used to rule out those configurations where only right-handed massless mode are trapped.

In addition, it is worthwhile to mention the role played by the parameter $k_{1,2}$ on the brane internal structure. As $k_{1,2}$ grows the brane undergoes a transition from a single into a two-brane. Therefore, the parameter $k_{1,2}$ can be regarded as phase transition parameter controlling the brane splitting process.



\section*{Acknowledgments}
\hspace{0.5cm}The authors thank the Conselho Nacional de Desenvolvimento Cient\'{\i}fico e Tecnol\'{o}gico (CNPq), grants n$\textsuperscript{\underline{\scriptsize o}}$ 312356/2017-0 (JEGS) and n$\textsuperscript{\underline{\scriptsize o}}$ 308638/2015-8 (CASA), and Coordena\c{c}\~{a}o de Aperfei\c{c}oamento do Pessoal de N\'{i}vel Superior (CAPES), for financial support. The authors also thank the anonymous referee for valuable comments and suggestions.


\begin{thebibliography}{99}

\bibitem{rs}
    L.~Randall and R.~Sundrum, Phys.\ Rev.\ Lett.\  {\bf 83} (1999) 4690. 
        
\bibitem{rs2}
    L.~Randall and R.~Sundrum, Phys.\ Rev.\ Lett.\  {\bf 83} (1999) 3370.

\bibitem{cosmologicalconstant}
J.~M.~Schwindt and C.~Wetterich,
  Nucl.\ Phys.\ B {\bf 726} (2005) 75.

\bibitem{darkmatter}
  T.~Gherghetta and B.~von Harling,
  JHEP {\bf 1004} (2010) 039.

\bibitem{Csaki1}   
    C.~Csaki, J.~Erlich, T.~J.~Hollowood and Y.~Shirman, Nucl.\ Phys.\ B {\bf 581} (2000) 309.

\bibitem{Rosa2020}
J.~L.~Rosa, D.~A.~Ferreira, D.~Bazeia and F.~S.~N.~Lobo,
Eur. Phys. J. C \textbf{81} (2021) 20.

\bibitem{Kehagias}
  A.~Kehagias and K.~Tamvakis,
Phys.\ Lett.\ B {\bf 504} (2001) 38.


\bibitem{Almeida2009}
C.~A.~S.~Almeida, M.~M.~Ferreira, Jr., A.~R.~Gomes and R.~Casana,
Phys. Rev. D \textbf{79} (2009) 125022.

\bibitem{Hayashi1979}
K.~Hayashi and T.~Shirafuji,
Phys. Rev. D \textbf{19} (1979) 3524.

\bibitem{deAndrade1997}
V.~C.~de Andrade and J.~G.~Pereira,
Phys. Rev. D \textbf{56} (1997) 4689.


\bibitem{andrade2000}
V. C. de Andrade, L. C. T. Guillen, and J. G. Pereira, Phys. Rev. Lett. \textbf{84} (2000) 4533.

\bibitem{ferraro2007} Rafael Ferraro and Franco Fiorini, Phys. Rev. D \textbf{75} (2007) 084031.

\bibitem{lobo2012} C. G. B\"{o}hmer, T. Harko, and Francisco S. N. Lobo, Phys. Rev. D \textbf{85} (2012) 044033.

\bibitem{Aldrovandi} R.~Aldrovandi and J.~G.~Pereira, Fundam. Theor. Phys. \textbf{173} (2013).

\bibitem{cai2016} Yi-Fu Cai, S. Capozziello, M. De Laurentis, and E. N. Saridakis, Rep. Prog. Phys. \textbf{79} (2016) 106901.

\bibitem{koi2020} J. Jim\'{e}nez, L. Heisenberg, D. Iosifidis, A. Jim\'{e}nez-Cano, and T. Koivisto, Phys. Lett. B \textbf{805} (2020) 135422

\bibitem{olmo2020} C. Bejarano, A. Delhom, A. Jim\'{e}nez-Cano, G. Olmo, and D. Rubiera-Garcia, Phys. Lett. B \textbf{802} (2020) 135275

\bibitem{RandjbarDaemi2000}
S.~Randjbar-Daemi and M.~E.~Shaposhnikov,
Phys. Lett. B \textbf{492} (2000) 361.

\bibitem{Liu2009}
Y.~X.~Liu, J.~Yang, Z.~H.~Zhao, C.~E.~Fu and Y.~S.~Duan,
Phys. Rev. D \textbf{80} (2009) 065019.

\bibitem{Liu2009a}
Y.~X.~Liu, H.~T.~Li, Z.~H.~Zhao, J.~X.~Li and J.~R.~Ren,
JHEP \textbf{10} (2009) 091.

\bibitem{Liu2008}
Y.~X.~Liu, L.~D.~Zhang, L.~J.~Zhang and Y.~S.~Duan,
Phys. Rev. D \textbf{78} (2008) 065025.



\bibitem{Liu2009b}
Y.~X.~Liu, C.~E.~Fu, L.~Zhao and Y.~S.~Duan,
Phys. Rev. D \textbf{80} (2009) 065020.

\bibitem{Liu2008b}
Y.~X.~Liu, L.~D.~Zhang, S.~W.~Wei and Y.~S.~Duan,
JHEP \textbf{08}(2008) 041.

\bibitem{Obukhov2002}
Y.~N.~Obukhov and J.~G.~Pereira,
Phys. Rev. D \textbf {67} (2003) 044016.



\bibitem{Ulhoa2016}
S.~C.~Ulhoa, A.~F.~Santos and F.~C.~Khanna,
Gen. Rel. Grav. {\bf 49} (2017) 54.

\bibitem{Dantas2013}
D.~M.~Dantas, J.~E.~G.~Silva and C.~A.~S.~Almeida,
Phys. Lett. B \textbf {725} (2013) 425.
  
\bibitem{Sousa2014}
L.~J.~S.~Sousa, C.~A.~S.~Silva, D.~M.~Dantas and C.~A.~S.~Almeida,
Phys. Lett. B \textbf {731} (2014) 64.
  
\bibitem{Dantas}
D.~M.~Dantas, D.~F.~S.~Veras, J.~E.~G.~Silva and C.~A.~S.~Almeida,
Phys. Rev. D {\bf 92} (2015) 104007.


\bibitem{Mitra2017}
J.~Mitra, T.~Paul and S.~SenGupta,
Eur. Phys. J. C \textbf {77} (2017) 833.

\bibitem{Buyukdag2018}
Y.~Buyukdag, T.~Gherghetta and A.~S.~Miller,
Phys. Rev. D \textbf {99} (2019) 035046. 

\bibitem{Wang2019}
L.~L.~Wang, H.~Guo, C.~E.~Fu and Q.~Y.~Xie,
 `` Gravity and Matters on a pure geometric thick polynomial f(R) brane '', \url{https://arxiv.org/abs/1912.01396}.  

\bibitem{Yang2012}
J.~Yang, Y.~-L.~Li, Y.~Zhong and Y.~Li,
  Phys.\ Rev.\ D {\bf 85} (2012) 084033.  
  
\bibitem{Yang2017}
K.~Yang, W.~D.~Guo, Z.~C.~Lin and Y.~X.~Liu,
Phys. Lett. B \textbf{782} (2018) 170.



\bibitem{Bahamonde2015}
S.~Bahamonde, C.~G.~B\"ohmer and M.~Wright,
Phys. Rev. D \textbf{92} (2015) 104042.

\bibitem{Wright2016}
M.~Wright,
Phys. Rev. D \textbf{93} (2016) 103002.

\bibitem{Bahamonde2016}
S.~Bahamonde and S.~Capozziello,
Eur. Phys. J. C \textbf{77} (2017) 107.

\bibitem{Franco2020}
G.~A.~R.~Franco, C.~Escamilla-Rivera and J.~Levi Said,
Eur. Phys. J. C \textbf{80} (2020) 677.



\bibitem{EscamillaRivera2019}
C.~Escamilla-Rivera and J.~Levi Said,
Class. Quant. Grav. \textbf{37} (2020) 165002.

\bibitem{Bahamonde2016a}
S.~Bahamonde, M.~Zubair and G.~Abbas,
Phys. Dark Univ. \textbf{19} (2018) 78.

\bibitem{Caruana2020}
M.~Caruana, G.~Farrugia and J.~Levi Said,
Eur. Phys. J. C \textbf{80} (2020) 640.

\bibitem{Pourbagher2020}
A.~Pourbagher and A.~Amani,
Mod. Phys. Lett. A \textbf{35} (2020) 2050166.

\bibitem{Bahamonde2020a}
S.~Bahamonde, V.~Gakis, S.~Kiorpelidi, T.~Koivisto, J.~Levi Said and E.~N.~Saridakis,
Eur. Phys. J. C \textbf{81} (2021) 53.

\bibitem{Azhar2020}
N.~Azhar, A.~Jawad and S.~Rani,
Phys. Dark Univ. \textbf{30} (2020) 100724.


\bibitem{Bhattacharjee2020}
S.~Bhattacharjee,
Phys. Dark Univ. \textbf{30} (2020) 100612.

\bibitem{Abedi2017}
H.~Abedi and S.~Capozziello,
Eur. Phys. J. C \textbf{78} (2018) 474.

\bibitem{Allan}
A.~R.~P.~Moreira, J.~E.~G.~Silva, F.~C.~E.~Lima and C.~A.~S.~Almeida,
 Phys. Rev. D \textbf{103} (2021) 064046.

\bibitem{BLi2010}
B.~Li, T.~P.~Sotiriou and J.~D.~Barrow,
Phys. Rev. D \textbf{83} (2011) 064035.

\bibitem{BLi2011}
B.~Li, T.~P.~Sotiriou and J.~D.~Barrow,
Phys. Rev. D \textbf{83} (2011) 104017.

\bibitem{Sotiriou2010}
T.~P.~Sotiriou, B.~Li and J.~D.~Barrow,
Phys. Rev. D \textbf{83} (2011) 104030.

\bibitem{Capozziello2019}
S.~Capozziello, M.~Capriolo and L.~Caso,
Eur. Phys. J. C \textbf{80} (2020) 156.
 

\bibitem{Gremm1999}
M.~Gremm,
Phys. Lett. B \textbf{478} (2000) 434.

\bibitem{Bazeia2007}
D.~Bazeia, A.~R.~Gomes and L.~Losano,
Int. J. Mod. Phys. A \textbf{24} (2009) 1135.

\bibitem{Andrade2001}
V.~C.~de Andrade, L.~C.~T.~Guillen and J.~G.~Pereira,
Phys. Rev. D \textbf {64} (2001) 027502.


 







  
\end{thebibliography}

\end{document}